\documentclass[12pt]{article}

\usepackage[utf8]{inputenc}
\usepackage[T1]{fontenc}
\usepackage{lmodern}
\usepackage{amsmath,amssymb}
\usepackage{amsthm}
\usepackage{dsfont}
\usepackage{mathrsfs}
\usepackage{float}
\usepackage{graphicx}
\usepackage{babel}
\usepackage{natbib}
\usepackage{setspace}
\theoremstyle{plain}
\newtheorem{assumption}{\protect\assumptionname}
\theoremstyle{plain}

\theoremstyle{definition}
 \newtheorem{example}{\protect\examplename}
\theoremstyle{remark}
\newtheorem{rem}{\protect\remarkname}
\theoremstyle{plain}

\theoremstyle{plain}
\newtheorem{thm}{\protect\theoremname}
\theoremstyle{plain}
\newtheorem{lem}{\protect\lemmaname}

\providecommand{\assumptionname}{Assumption}
\providecommand{\corollaryname}{Corollary}
\providecommand{\examplename}{Example}
\providecommand{\propositionname}{Proposition}
\providecommand{\remarkname}{Remark}
\providecommand{\theoremname}{Theorem}
\providecommand{\lemmaname}{Lemma}

\usepackage[top=1in, bottom=1in, left=1in, right=1in]{geometry}

\title{Policy Targeting with Market Equilibrium\thanks{I would like to thank Toru Kitagawa, Susanne Schennach, Andriy Norets, Soonwoo Kwon, Jonathan Roth, and Peter Hull for helpful comments. I gratefully acknowledge financial support from Department of Economics, Brown University (Merit Dissertation Fellowship).}}
\author{Gyungbae Park\thanks{School of Management and Economics, The Chinese University of Hong Kong, Shenzhen}}
\date{September 16th, 2026}

\begin{document}
\maketitle

%% Abstract
\begin{abstract}
%% Text of abstract
This paper develops a framework for individualized treatment allocation when interventions shift equilibrium prices and generate spillovers across treated and untreated units. The planner chooses which units receive a subsidy while allowing equilibrium prices to adjust endogenously. We show that the resulting welfare function is supermodular under broad and interpretable conditions, implying complementarity across treatment assignments and enabling exact polynomial-time optimization. This structure clarifies how equilibrium spillovers shape the trade-off between universal and targeted distribution and makes the planner's problem computationally tractable despite interactions across units. We characterize when universal or targeted subsidies are optimal and show how market conditions and heterogeneity shape the optimal allocation. We further establish statistical guarantees for plug-in allocation under estimation uncertainty in demand and supply. Finally, we illustrate the framework in a coupon allocation problem calibrated with household expenditure data from the Philippines.
\end{abstract}

\vspace{1em}

\section{Introduction}
\label{s1}
%% Labels are used to cross-reference an item using \ref command.

Allocating scarce policy resources across heterogeneous populations is a central problem in economics and public policy. Planners must often decide whether to deliver interventions universally or to target specific groups. A common form of such intervention is a selective subsidy that lowers the price of one good while leaving other goods at market prices. Examples include discount coupons for healthy foods, incentives for green-energy products, or housing vouchers that apply only to particular sectors. These policies are motivated by social objectives such as improving nutrition, promoting environmental sustainability, or increasing access to essential goods. Yet, by changing the relative prices of the subsidized good, they also reshape equilibrium outcomes. This asymmetry raises a fundamental question: when is it welfare-improving to subsidize one good but not another, and how should such subsidies be distributed across heterogeneous groups? When subsidies expand demand for the targeted good, heterogeneous demand responses across groups shift equilibrium prices and alter the welfare of both treated and untreated households. Designing allocation rules that maximize social welfare in the presence of such equilibrium price adjustments is both practically important and conceptually challenging.

This paper provides a framework for treatment allocation in markets where policy interventions affect outcomes through equilibrium price spillovers. While the framework applies broadly to subsidies, vouchers, or transfers, a leading example is the use of discount coupons for a subsidized good. Coupon programs are appealing because they can stimulate demand quickly, yet their distribution raises fundamental design questions: should they be issued universally or targeted to specific groups, and how do the resulting price adjustments affect the welfare of both recipients and non-recipients? By explicitly modeling these market-wide spillovers, the framework highlights when universal distribution is optimal, when targeted allocation dominates, and how the planner's problem of who to target can be solved tractably.

Existing research on policy targeting and treatment choice has developed powerful methods for learning optimal allocation rules, but most of this literature assumes no spillovers across units. The standard framework, dating back to the formulation of statistical treatment rules (\citet{Manski2004}) and extended by \citet{Hirano.and.Porter.2009}, \citet{Kitagawa.and.Tetenov2018}, and \citet{Athey.and.Wager2021}, evaluates welfare under the stable-unit-treatment-value assumption (SUTVA) (\citet{Rubin1980}), which requires that each individual's outcome depends only on their own treatment and not on others' treatments. Foundational contributions formalize interference and partial interference (\citet{Hudgens.and.Halloran.2008}; \citet{Aronow.and.Samii.2017}). More recent works on treatment choice relax the SUTVA assumption by allowing for interference through networks (\citet{Viviano2024}; \citet{Kitagawa.and.Wang.2023}; \citet{Ananth2021}), strategic interactions (\citet{Sahoo.and.Wager.2025}; \citet{Kitagawa.and.Wang.2026}), displacement effects of job training program (\citet{Crépon.et.al.2013}), discrete choice models with social interactions (\citet{Bhattacharya.et.al.2024}), or market equilibrium effects (\citet{Munro.et.al.2025}). These contributions demonstrate that ignoring spillovers can lead to misleading policy recommendations.

Nonetheless, existing approaches remain limited in important ways. Many restrict the class of admissible policies in order to control complexity, while others focus on equilibrium-stable rules that prevent prices from adjusting. For example, \citet{Munro.et.al.2025} studies treatment effects when prices are endogenous but concentrate on equilibrium-stable targeting rules designed not to alter market prices. Such restrictions make the analysis tractable but may fall short of maximizing social welfare and are often motivated by the desire to avoid estimating price elasticities. A key open challenge is to develop a framework that allows interventions to reshape market outcomes while still enabling tractable welfare analysis and implementable allocation rules.

In our framework, the policymaker's welfare function aggregates individual utilities across groups, capturing both the direct benefits of treatment and the indirect effects that arise as equilibrium prices adjust to market-wide changes in demand. The decision to treat one group of identical consumers affects market prices, which in turn influence the outcomes of all groups. Each group is represented by a single representative consumer. Hence, the optimal allocation must be evaluated jointly at the market level. A brute-force enumeration of all possible treatment assignments is computationally infeasible, as the number of allocations grows exponentially with the number of groups. To address this, we establish conditions under which the planner's objective function exhibits supermodularity, meaning that the benefit of treating one group is larger when others are also treated. This property is important both computationally and economically. Computationally, it ensures that the planner's problem, otherwise intractable, can be solved exactly in polynomial time using well-established algorithms in combinatorial optimization (\citet{Schrijver2000}; \citet{Iwata.et.al.2001}; \citet{Orlin2009}). Economically, it highlights that treatments are complementary through equilibrium price adjustments, yet the welfare function may remain non-monotonic in treatment intensity. Thus, policy evaluation must account for how interventions interact across groups.

Our analysis makes two main contributions. First, we show that the policymaker's welfare function is supermodular in treatment assignments under broad and interpretable conditions. Supermodularity implies that treatments are complements, so the welfare gain from treating an additional group is larger when others are also treated. This complementarity structure not only clarifies when universal versus targeted distribution is optimal, but also ensures tractability, since supermodular maximization can be solved exactly in polynomial time using established algorithms. We establish supermodularity in a Cobb-Douglas benchmark and show that it extends under broad and interpretable conditions, including convex indirect utility, dominance of own-price responses, bounded relative risk aversion, and submodular equilibrium price responses. These results demonstrate that the tractability and comparative statics are not confined to special functional forms, but hold more generally.

Second, we develop a framework for treatment allocation in markets with equilibrium price spillovers that delivers both tractability and statistical guarantees. The framework allows interventions to reshape market outcomes while still permitting welfare evaluation. This contribution is important because much of the existing literature focuses on equilibrium-stable targeting, where the planner optimizes welfare only among allocations that keep equilibrium prices fixed. Such a restriction can overlook policies that would induce beneficial price adjustments. In contrast, our framework allows the planner to consider allocations that endogenously shift equilibrium prices, capturing both the direct effect of treatment and the indirect effect operating through market feedback. This generalization expands the scope of feasible policies while maintaining theoretical and computational tractability.

In addition, we present an empirical illustration to demonstrate how the framework can be applied in practice. Using moments of household expenditure data from \citet{Filmer.et.al.2023}, we calibrate key parameters, including supply elasticities, and group-level heterogeneity, and use them to evaluate welfare under different allocation rules. This exercise connects the planner’s welfare function to real-world data and shows how the method can guide policy choices.

This paper also relates to the growing literature on fiscal stimulus using consumer-financed mechanisms such as digital coupons. Recent empirical studies (\citet{Liu.et.al.2021}; \citet{Xing.et.al.2023}; \citet{Ding.et.al.2025}; \citet{Chen.et.al.2026}) document substantial short-run spending multipliers and highlight distributional concerns. Our contribution is complementary: rather than providing additional empirical estimates, we develop a theoretical framework that incorporates general equilibrium price adjustments into the analysis of coupon-based programs. This framework delivers clear conditions under which universal versus targeted subsidies should be employed, offering normative guidance that complements the positive evidence from this empirical literature.

More broadly, our work contributes to the literature on supermodularity in economics. Supermodularity is central to lattice-theoretic methods and monotone comparative statics (\citet{Topkis.1978}; \citet{Milgrom.and.Shannon.1994}). It underlies influential applications in games of strategic interaction (\citet{Milgrom.and.Roberts.1990}; \citet{Vives1990}; \citet{Vives2005}), industrial organization and auctions (\citet{Milgrom2000}), and macroeconomic models with strategic complementarities (\citet{Cooper.and.John.1988}). These contributions show how complementarity structures yield robust comparative statics and tractable equilibria. In the present context, the complementarity structure of the planner's welfare function offers a distinctive theoretical and computational contribution, yielding both tractable algorithms and clear policy implications. Our findings thus link the policy-learning literature to a broader tradition in economics that exploits complementarity structures to obtain sharp and interpretable results.

Beyond discount coupons, the framework applies broadly to interventions that raise demand in markets with finite supply, such as housing vouchers (\citet{Jacob.and.Ludwig.2012}), tuition subsidies (\citet{Dynarski2003}), or health-product coupons (\citet{Dupas2014}). By incorporating market-equilibrium responses into the policy learning framework, we provide theoretical foundations for when universal subsidies are optimal, when targeted transfers dominate, and how structural features of demand and supply shape the tractability of optimal allocation.

The remainder of the paper is organized as follows. Section~\ref{s2} presents the theoretical framework and develops the main results. Section~\ref{s3} provides an empirical illustration that calibrates key parameters using household expenditure data. Section~\ref{s4} concludes. Proofs and additional derivations are collected in the Appendix.

\section{Model}
\label{s2}

We study a market with $G$ groups, indexed by $g=1,2,\ldots,G.$ Each group consists of identical consumers with the same preferences and income, and is represented by a single representative consumer. Two goods are available in the market: good $\alpha$ with price $p_{\alpha}$ and good $\beta$ with price $p_{\beta}.$ The policy intervention is a binary treatment $W_{g}\in\left\{ 0,1\right\} $ for each group, and let $\boldsymbol{W}=\left(W_{1},W_{2},\ldots,W_{G}\right)^{'}$ denote the vector of assignments. Treated groups receive a $\left(100\times\tau_{\alpha}\right)\%$ discount coupon for good $\alpha,$ where $0<\tau_{\alpha}<1.$ This asymmetry captures policies that subsidize one good for public health or environmental reasons, such as discount coupons for healthy foods or incentives for green-energy products, while the price of good $\beta$ remains market-determined.

Given a price vector $p=\left(p_{\alpha},p_{\beta}\right)^{'}\in\mathbb{R}_{++}^{2},$ consumers in group $g$ choose a consumption bundle $Q^{\left(g\right)}=\left(Q_{\alpha}^{\left(g\right)},Q_{\beta}^{\left(g\right)}\right)^{'}$ by solving
\begin{equation*}
    \max_{Q^{\left(g\right)}}U^{\left(g\right)}\left(Q_{\alpha}^{\left(g\right)},Q_{\beta}^{\left(g\right)}\right)\;\mathrm{s}.\mathrm{t}.\;\left(1-\tau_{\alpha}W_{g}\right)p_{\alpha}Q_{\alpha}^{\left(g\right)}+p_{\beta}Q_{\beta}^{\left(g\right)}\leq M_{g},
\end{equation*}
where $M_{g}>0$ is income and $U^{\left(g\right)}\left(\cdot\right)$ is strictly concave and twice differentiable. It is convenient to define the coupon-adjusted price for group $g$ as
\begin{equation*}
    \tilde{p}_{\alpha,g}\equiv\left(1-\tau_{\alpha}W_{g}\right)p_{\alpha}.
\end{equation*}
The corresponding demand functions can then be expressed as
\begin{eqnarray*}
    Q_{\alpha}^{\left(g\right)} & = & Q_{\alpha}^{\left(g\right)}\left(\tilde{p}_{\alpha,g},p_{\beta}\right)=Q_{\alpha}^{\left(g\right)}\left(\left(1-\tau_{\alpha}W_{g}\right)p_{\alpha},p_{\beta}\right)\\
    Q_{\beta}^{\left(g\right)} & = & Q_{\beta}^{\left(g\right)}\left(\tilde{p}_{\alpha,g},p_{\beta}\right)=Q_{\beta}^{\left(g\right)}\left(\left(1-\tau_{\alpha}W_{g}\right)p_{\alpha},p_{\beta}\right).
\end{eqnarray*}
A key feature of the model is that treatment influences behavior only through the coupon-adjusted price. In other words, the policy does
not alter preferences or directly change incomes; it simply lowers the effective price of good $\alpha.$

Aggregate supply is given by strictly convex, monotonically increasing functions $S_{\alpha}\left(p_{\alpha}\right)$ and $S_{\beta}\left(p_{\beta}\right).$ Market clearing requires
\begin{equation*}
    \left[\begin{array}{c}
\sum_{g\in\left\{ 1,2,\cdots,G\right\} }Q_{\alpha}^{\left(g\right)}\left(\left(1-\tau_{\alpha}W_{g}\right)p_{\alpha},p_{\beta}\right)-S_{\alpha}\left(p_{\alpha}\right)\\
\sum_{g\in\left\{ 1,2,\cdots,G\right\} }Q_{\beta}^{\left(g\right)}\left(\left(1-\tau_{\alpha}W_{g}\right)p_{\alpha},p_{\beta}\right)-S_{\beta}\left(p_{\beta}\right)
\end{array}\right]=0.
\end{equation*}
The equilibrium price vector is denoted $p\left(\boldsymbol{W}\right)=\left(p_{\alpha}\left(\boldsymbol{W}\right),p_{\beta}\left(\boldsymbol{W}\right)\right)^{'}.$ The indirect utility of group $g$ is
\begin{eqnarray*}
     &  & V^{\left(g\right)}\left(\boldsymbol{W}\right)\\
      & \equiv & U^{\left(g\right)}\left(Q_{\alpha}^{\left(g\right)}\left(\tilde{p}_{\alpha,g}\left(\boldsymbol{W}\right),p_{\beta}\left(\boldsymbol{W}\right)\right),Q_{\beta}^{\left(g\right)}\left(\tilde{p}_{\alpha,g}\left(\boldsymbol{W}\right),p_{\beta}\left(\boldsymbol{W}\right)\right)\right)\\
      & = & U^{\left(g\right)}\left(Q_{\alpha}^{\left(g\right)}\left(\left(1-\tau_{\alpha}W_{g}\right)p_{\alpha}\left(\boldsymbol{W}\right),p_{\beta}\left(\boldsymbol{W}\right)\right),Q_{\beta}^{\left(g\right)}\left(\left(1-\tau_{\alpha}W_{g}\right)p_{\alpha}\left(\boldsymbol{W}\right),p_{\beta}\left(\boldsymbol{W}\right)\right)\right),
\end{eqnarray*}
which depends not only on its own treatment but also on others' treatments through equilibrium prices.

For econometric interpretation, it is useful to rewrite the model using potential outcomes. Let $Y_{g}\left(\boldsymbol{W}\right)$
denote the potential outcome of the representative consumer in group $g$ under treatment assignment vector $\boldsymbol{W}.$ Then $Y_{g}\left(\boldsymbol{W}\right)=V^{\left(g\right)}\left(\boldsymbol{W}\right),$ where $V^{\left(g\right)}\left(\boldsymbol{W}\right)$ denotes the indirect utility of the representative consumer in group $g.$ In this framework, the treatment vector $\boldsymbol{W}$ affects outcomes only through the equilibrium price vector, which serves as a low-dimensional exposure mapping in the sense of \citet{Aronow.and.Samii.2017}. The observed outcome is $Y_{g}=Y_{g}\left(\boldsymbol{W}\right).$ The indirect utility is observable because, when all consumers within a group share identical preferences and income, their aggregate demand can be rationalized by a representative consumer whose indirect-utility function depends only on equilibrium prices and aggregate income. The SUTVA is replaced here by market-equilibrium interference, since $Y_{g}\left(\boldsymbol{W}\right)$ depends on the entire vector $\boldsymbol{W}$ through the equilibrium prices $p\left(\boldsymbol{W}\right).$ This notation makes explicit that we are in a potential-outcome framework with general interference.

Social welfare\footnote{One could alternatively define welfare to include non-market outcomes such as health status when good $\alpha$ represents a healthy product. However, incorporating such dimensions would require modeling how health evolves with consumption and interacts with market equilibrium. The resulting welfare function may fail to be supermodular in treatment assignments, in which case the planner's problem would lose the tractable lattice structure that underlies the current analysis.} is defined as
\begin{equation*}
    V\left(\boldsymbol{W}\right)=\sum_{g=1}^{G}\lambda_{g}Y_{g}\left(\boldsymbol{W}\right),
\end{equation*}
with the group weight $\lambda_{g}>0.$ This is a purely utilitarian social welfare function. \citet{Mas-Colell.et.al.1995} provides an interpretation of such a function as the expected utility of a representative individual 'behind the veil of ignorance'. More fundamentally, \citet{Harsanyi1955} shows that social welfare can be represented as a weighted sum of individual utilities.

In classic policy-learning problems, it is often assumed that $Y_{g}\left(\boldsymbol{W}\right)=Y_{g}\left(W_{g}\right)$ (no interference). Here, by contrast, $Y_{g}\left(\boldsymbol{W}\right)$ depends on all $\boldsymbol{W}_{-g}$ via equilibrium prices, representing a specific structured form of interference. The policymaker's problem is
\begin{equation*}
    \max_{\boldsymbol{W}\in\left\{ 0,1\right\} ^{G}}V\left(\boldsymbol{W}\right).
\end{equation*}
Because the number of possible allocations grows exponentially with $G,$ a brute-force search over all $2^{G}$ assignments is computationally infeasible for large $G.$ If, however, the welfare function $V\left(\boldsymbol{W}\right)$ is supermodular, then the optimal allocation can be recovered in exact polynomial time using well-established algorithms (\citet{Schrijver2000}; \citet{Iwata.et.al.2001}; \citet{Orlin2009}).

To analyze the planner's problem, it is useful to introduce a structural property that will play a central role in the analysis. The planner's welfare function $V\left(\boldsymbol{W}\right)$ is supermodular if, for any two allocations $\boldsymbol{W},\boldsymbol{W}^{'},$
\begin{equation*}
    V\left(\boldsymbol{W}\right)+V\left(\boldsymbol{W}^{'}\right)\leq V\left(\boldsymbol{W}\wedge\boldsymbol{W}^{'}\right)+V\left(\boldsymbol{W}\vee\boldsymbol{W}^{'}\right),
\end{equation*}
where $\left(\boldsymbol{W}\wedge\boldsymbol{W}^{'}\right)_{g}=\min\left\{ W_{g},W_{g}^{'}\right\} $ and $\left(\boldsymbol{W}\vee\boldsymbol{W}^{'}\right)_{g}=\max\left\{ W_{g},W_{g}^{'}\right\} .$ Intuitively, this means that treatment assignments are complementary: the welfare gain from treating one group is larger when other groups are also treated because equilibrium price adjustments amplify the joint benefits of interventions. However, complementarity does not imply that welfare is monotonic in the number of treated groups. As more groups are treated, higher aggregate demand may raise prices and potentially reduce welfare, so the overall welfare function can remain non-monotonic despite being supermodular. Moreover, when $V\left(\boldsymbol{W}\right)$ admits a smooth extension to the hypercube $\left[0,1\right]^{G},$ supermodularity is equivalent to nonnegative cross-partial derivatives,
\begin{equation*}
    \frac{\partial^{2}V}{\partial W_{s}\partial W_{r}}\geq0\quad\mathrm{for}\;\mathrm{all}\;r\neq s,
\end{equation*}
so that spillover effects reinforce each other across groups. This property provides both economic intuition and tractability for the planner's optimization problem.

\begin{rem}\label{re1}
A policymaker may be concerned about the cost of issuing discount coupons, since in practice the total number of coupons is often limited by the government's budget constraint. In this scenario, the policymaker can consider the following optimization problem:
\begin{equation*}
    \max_{\boldsymbol{W}\in\left\{ 0,1\right\} ^{G}}V\left(\boldsymbol{W}\right)-\xi\sum_{g=1}^{G}W_{g}
\end{equation*}
for $\xi\geq0.$ This formulation incorporates the cost of issuing discount coupons through the weight parameter $\xi.$ Because the linear function is modular (that is, both supermodular and submodular), the objective function remains supermodular as long as $V\left(\boldsymbol{W}\right)$ is supermodular. Therefore, the cost-adjusted optimization problem can be solved using the same off-the-shelf algorithm as before. When $\xi$ is large, the policymaker becomes more reluctant to issue discount coupons, leading to fewer treated groups.
\end{rem}

\subsection{Cobb-Douglas Utilities}

To build intuition, we begin with a benchmark in which preferences are Cobb-Douglas and supply is constant-elasticity. This benchmark not only yields closed-form solutions but also ensures that the welfare function is supermodular.

\begin{assumption}\label{as1}
    Each group $g$ has Cobb-Douglas utility
    \begin{equation*}
        U^{\left(g\right)}\left(Q_{\alpha}^{\left(g\right)},Q_{\beta}^{\left(g\right)}\right)	=	\gamma_{g}\log Q_{\alpha}^{\left(g\right)}+\left(1-\gamma_{g}\right)\log\left(Q_{\beta}^{\left(g\right)}\right),\qquad0<\gamma_{g}<1,
    \end{equation*}
    subject to the budget constraint
    \begin{equation*}
        \left(1-\tau_{\alpha}W_{g}\right)p_{\alpha}Q_{\alpha}^{\left(g\right)}+p_{\beta}Q_{\beta}^{\left(g\right)}\leq M_{g}.
    \end{equation*}
    Aggregate supplies are given by
    \begin{equation*}
        \left[\begin{array}{c}
\log S_{\alpha}\\
\log S_{\beta}
\end{array}\right]=\left[\begin{array}{c}
\log\delta_{\alpha}+\kappa_{\alpha}\log p_{\alpha}\\
\log\delta_{\beta}+\kappa_{\beta}\log p_{\beta}
\end{array}\right]
    \end{equation*}
    where $\kappa_{\alpha},\kappa_{\beta}>0.$
\end{assumption}

This specification is tractable and informative. Cobb-Douglas preferences generate simple demand functions, while constant-elasticity supply links price responses directly to elasticity parameters. Together, they provide a clean environment to study welfare effects of coupons and to establish structural properties such as supermodularity.

\begin{thm}\label{th1}
Suppose that Assumption \ref{as1} holds. Then, for any given parameters $\gamma_{g},$ $M_{g},$ $\lambda_{g},$ and $\kappa_{\alpha},$ the policymaker's value function $V\left(\boldsymbol{W}\right)$ is supermodular in $\boldsymbol{W}.$
\end{thm}

Theorem \ref{th1} establishes that treatment assignments are complements: the welfare gain from treating an additional group is larger when more groups are already treated. Computationally, this makes optimization feasible in polynomial time, while economically, it highlights that coupon subsidies interact in a complementary way across groups through equilibrium price adjustments. Welfare, however, is not necessarily monotonic in treatment intensity, as spillover effects may be positive or negative depending on price responses.

\subsection{Illustrative Examples with Cobb-Douglas Utilities}

Theorem \ref{th1} also provides a practical strategy. Instead of enumerating all allocations, the policymaker can apply existing supermodular maximization algorithms. The qualitative implications are especially transparent in the two-group case, where closed-form welfare expressions are concise:
\begin{eqnarray*}
    V\left(0,0\right) & = & -\left\{ \frac{\log\left(\gamma_{1}M_{1}+\gamma_{2}M_{2}\right)}{1+\kappa_{\alpha}}\left(\gamma_{1}+\gamma_{2}\right)\right\} \\
    V\left(1,0\right) & = & -\gamma_{1}\log\left(1-\tau_{\alpha}\right)-\frac{\log\left(\frac{\gamma_{1}M_{1}}{1-\tau_{\alpha}}+\gamma_{2}M_{2}\right)}{1+\kappa_{\alpha}}\left(\gamma_{1}+\gamma_{2}\right)\\
    V\left(0,1\right) & = & -\gamma_{2}\log\left(1-\tau_{\alpha}\right)-\frac{\log\left(\gamma_{1}M_{1}+\frac{\gamma_{2}M_{2}}{1-\tau_{\alpha}}\right)}{1+\kappa_{\alpha}}\left(\gamma_{1}+\gamma_{2}\right)\\
    V\left(1,1\right) & = & -\left(\gamma_{1}+\gamma_{2}\right)\log\left(1-\tau_{\alpha}\right)-\frac{\log\left(\frac{\gamma_{1}M_{1}}{1-\tau_{\alpha}}+\frac{\gamma_{2}M_{2}}{1-\tau_{\alpha}}\right)}{1+\kappa_{\alpha}}\left(\gamma_{1}+\gamma_{2}\right).
\end{eqnarray*}
These closed-form welfare expressions are useful for deriving the comparative statics in Examples \ref{ex1}-\ref{ex5}, since they allow direct comparison of allocations as functions of incomes, preferences, and supply elasticity. For simplicity, throughout the examples, assume $\lambda_{g}=1$ (equal group weight) and $\xi=0$ (no cost of issuing coupons).

\begin{example}\label{ex1}
$V\left(1,1\right)>V\left(0,0\right)$.
\end{example}
This result shows that universal treatment always improves welfare relative to providing no treatment. The intuition is straightforward: issuing coupons raises effective demand, and because all groups benefit from the resulting price adjustments, aggregate welfare strictly increases. Even when supply is not perfectly elastic, some of the benefits of treatment diffuse to untreated groups through spillover effects, ensuring that total welfare does not fall. As a result, doing nothing is never optimal unless the cost of issuing coupons is taken into account. From this point onward, the planner only needs to compare three candidates: treating only the first group, treating only the second group, or treating both groups universally.

\begin{example}\label{ex2}
When $\gamma_{1}=\gamma_{2}=\gamma,$ then
\begin{equation*}
    V\left(1,0\right)\underset{<}{\overset{>}{=}}V\left(0,1\right)\quad\mathrm{iff}\quad M_{1}\underset{>}{\overset{<}{=}}M_{2}.
\end{equation*}
\end{example}
With identical preferences across groups, the only heterogeneity comes from income. Coupons expand effective demand proportionally to income, but the welfare gain is larger when resources are scarce, since the marginal utility of additional consumption is higher for poorer groups. This means that, when preferences are homogeneous, the planner maximizes welfare by targeting the group with lower income. This result clarifies under what conditions treating only the first group is better than treating only the second group.

\begin{example}\label{ex3}
When $M_{1}=M_{2}=M,$ then
\begin{equation*}
    V\left(1,0\right)\underset{<}{\overset{>}{=}}V\left(0,1\right)\quad\mathrm{iff}\quad\gamma_{1}\underset{<}{\overset{>}{=}}\gamma_{2}
\end{equation*}
\end{example}
When incomes are equal, the relevant heterogeneity lies in preferences. A group that places a higher weight $\gamma$ on good $\alpha$ gains more from the coupon, since the subsidy directly lowers the effective price of the good they value most. This means the planner should prioritize the group with stronger demand for the subsidized good. Together with Example \ref{ex2}, this result shows that once no treatment has been ruled out, the choice between treating only the first or only the second group depends on income and preference heterogeneity. Because these comparisons are symmetric, the analysis can now focus without loss of generality on comparing treatment of the first group with universal treatment.

\begin{example}\label{ex4}
Suppose $\gamma=\gamma_{1}=\gamma_{2}.$ If $\kappa_{\alpha}\geq1,$ then $V\left(1,1\right)>V\left(1,0\right),$ and $V\left(1,1\right)>V\left(0,1\right).$
\end{example}
When supply is sufficiently elastic, expanding treatment to all groups is always welfare-improving. Issuing more coupons raises demand, but the resulting increase is absorbed mainly through higher quantities rather than higher prices. This means that universal treatment strictly dominates any targeted policy. In this case, the planner learns that if supply elasticity is high enough, the optimal allocation is unambiguously to treat everyone.

\begin{example}\label{ex5}
Suppose $\gamma=\gamma_{1}=\gamma_{2}.$ If $\kappa_{\alpha}<1$ and
$M_{1}<M_{2},$ then $V\left(1,1\right)<V\left(1,0\right)$ if and
only if $\frac{M_{1}}{M_{2}}<\frac{\left(1-\tau_{\alpha}\right)^{\frac{1+\kappa_{\alpha}}{2}}-\left(1-\tau_{\alpha}\right)}{1-\left(1-\tau_{\alpha}\right)^{\frac{1+\kappa_{\alpha}}{2}}}\approx\frac{1-\kappa_{\alpha}}{1+\kappa_{\alpha}}.$
\end{example}
When supply is inelastic, expanding treatment raises prices substantially. In this case, although issuing coupons still raises demand, the strong upward pressure on prices erodes some of the welfare gains. The condition derived in this example shows precisely when it is better to treat only the poorer group rather than to treat everyone. The cutoff is given by the inequality
\begin{equation*}
    M_{12}<\frac{1-\kappa_{\alpha}}{1+\kappa_{\alpha}}.
\end{equation*}
This means that universal treatment is dominated by targeted treatment exactly when the relative income of the poorer group is below this threshold, which depends inversely on the elasticity of supply. The sharper the inelasticity (the smaller $\kappa_{\alpha}$), the higher the threshold, and the more likely it is that targeting only the poorer group is optimal.

Taken together, these examples narrow the planner's problem step by step. Example \ref{ex1} rules out no treatment, since issuing coupons always raises demand and improves welfare. Examples \ref{ex2} and \ref{ex3} then determine whether it is better to treat only the first or only the second group, depending on whether income or preference heterogeneity dominates. Because those cases are symmetric, the analysis then focuses on comparing a targeted allocation with universal treatment. Examples \ref{ex4} and \ref{ex5} show that this final choice depends critically on supply conditions: when supply is elastic, universal treatment is always optimal, while when supply is inelastic, universal treatment may be dominated by targeting the poorer group. In particular, Example \ref{ex5} provides a sharp cutoff condition, showing that targeting is optimal when the income ratio of the poorer to the richer group falls below $\frac{1-\kappa_{\alpha}}{1+\kappa_{\alpha}}.$ These rich insights arise precisely because the analysis allows for equilibrium price adjustments; they would not be available if the planner restricted attention to price-stable allocations.

\subsection{Supermodularity}

It is also natural to ask whether the supermodularity of the welfare function is merely a special property of Cobb-Douglas preferences.
The following lemma decomposes the second-order mixed partial derivative of the welfare function, providing a route toward more general conditions. For simplicity, we focus on the case with equal group weights ($\lambda_{g}=1$), while noting that the supermodularity result extends straightforwardly to unequal weights. Recall that, for general indirect utilities $V^{\left(g\right)},$ the planner's welfare function is given by
\begin{eqnarray*}
     &  & V\left(\boldsymbol{W}\right)\\
     & = & \sum_{g=1}^{G}Y_{g}\left(\boldsymbol{W}\right)\\
     & = & \sum_{g=1}^{G}V^{\left(g\right)}\left(Q_{\alpha}^{\left(g\right)}\left(\left(1-\tau_{\alpha}W_{g}\right)p_{\alpha}\left(\boldsymbol{W}\right),p_{\beta}\left(\boldsymbol{W}\right)\right),Q_{\beta}^{\left(g\right)}\left(\left(1-\tau_{\alpha}W_{g}\right)p_{\alpha}\left(\boldsymbol{W}\right),p_{\beta}\left(\boldsymbol{W}\right)\right)\right).
\end{eqnarray*}
The following Lemma \ref{le2} decomposes the planner's welfare function.

\begin{lem}\label{le2}
For any $r\neq s,$
\begin{eqnarray*}
    \frac{\partial^{2}V}{\partial W_{s}\partial W_{r}} & = & \sum_{g=1}^{G}J_{s}^{\left(g\right)T}H^{\left(g\right)}J_{r}^{\left(g\right)}\\
    &  & -\tau_{\alpha}\left(\frac{\partial p_{\alpha}}{\partial W_{s}}\frac{\partial V^{\left(r\right)}}{\partial\tilde{p}_{\alpha,r}}+\frac{\partial p_{\alpha}}{\partial W_{r}}\frac{\partial V^{\left(s\right)}}{\partial\tilde{p}_{\alpha,s}}\right)\\
    &  & +\frac{\partial^{2}p_{\alpha}}{\partial W_{s}\partial W_{r}}\sum_{g=1}^{G}\left[\frac{\partial V^{\left(g\right)}}{\partial\tilde{p}_{\alpha,g}}\left(1-\tau_{\alpha}W_{g}\right)\right]\\
    &  & +\frac{\partial^{2}p_{\beta}}{\partial W_{s}\partial W_{r}}\sum_{g=1}^{G}\frac{\partial V^{\left(g\right)}}{\partial p_{\beta}}
\end{eqnarray*}
where
\begin{eqnarray*}
    J_{s}^{\left(g\right)} & \equiv & \left[\begin{array}{c}
\frac{\partial\tilde{p}_{\alpha,g}}{\partial W_{s}}\\
\frac{\partial p_{\beta}}{\partial W_{s}}
\end{array}\right]\\
    J_{r}^{\left(g\right)} & \equiv & \left[\begin{array}{c}
\frac{\partial\tilde{p}_{\alpha,g}}{\partial W_{r}}\\
\frac{\partial p_{\beta}}{\partial W_{r}}
\end{array}\right]\\
    H^{\left(g\right)} & \equiv & \left[\begin{array}{cc}
\frac{\partial^{2}V^{\left(g\right)}}{\partial\tilde{p}_{\alpha,g}^{2}} & \frac{\partial^{2}V^{\left(g\right)}}{\partial p_{\beta}\partial\tilde{p}_{\alpha,g}}\\
\frac{\partial^{2}V^{\left(g\right)}}{\partial\tilde{p}_{\alpha,g}\partial p_{\beta}} & \frac{\partial^{2}V^{\left(g\right)}}{\partial p_{\beta}^{2}}
\end{array}\right].
\end{eqnarray*}
Moreover, if $\frac{\partial Q_{\beta}^{\left(g\right)}}{\partial p_{\alpha}}=0$ for all $g,$ then
\begin{eqnarray*}
    &  & \frac{\partial^{2}V}{\partial W_{s}\partial W_{r}}\\
    & = & \frac{\partial p_{\alpha}}{\partial W_{s}}\frac{\partial p_{\alpha}}{\partial W_{r}}\sum_{g=1}^{G}\left(1-\tau_{\alpha}W_{g}\right)^{2}\frac{\partial^{2}V^{\left(g\right)}}{\partial\tilde{p}_{\alpha,g}^{2}}\\
    &  & -\tau_{\alpha}\left[\frac{\partial p_{\alpha}}{\partial W_{s}}\left(1-\frac{-\tilde{p}_{\alpha,r}\frac{\partial^{2}V^{\left(r\right)}}{\partial\tilde{p}_{\alpha,r}^{2}}}{\frac{\partial V^{\left(r\right)}}{\partial\tilde{p}_{\alpha,r}}}\right)\frac{\partial V^{\left(r\right)}}{\partial\tilde{p}_{\alpha,r}}+\frac{\partial p_{\alpha}}{\partial W_{r}}\left(1-\frac{-\tilde{p}_{\alpha,s}\frac{\partial^{2}V^{\left(s\right)}}{\partial\tilde{p}_{\alpha,s}^{2}}}{\frac{\partial V^{\left(s\right)}}{\partial\tilde{p}_{\alpha,s}}}\right)\frac{\partial V^{\left(s\right)}}{\partial\tilde{p}_{\alpha,s}}\right]\\
    &  & +\frac{\partial^{2}p_{\alpha}}{\partial W_{s}\partial W_{r}}\sum_{g=1}^{G}\left[\frac{\partial V^{\left(g\right)}}{\partial\tilde{p}_{\alpha,g}}\left(1-\tau_{\alpha}W_{g}\right)\right].
\end{eqnarray*}
\end{lem}
For supermodularity, it is required that $\frac{\partial^{2}V}{\partial W_{s}\partial W_{r}}>0$ for all distinct $r\neq s.$ To characterize sufficient conditions for positiveness, we now impose a sequence of structural assumptions. For simplicity, we focus on the case where $\frac{\partial Q_{\beta}^{\left(g\right)}}{\partial p_{\alpha}}=0.$

\begin{assumption}\label{as2}
    For all $g,$ the indirect utility function $V^{\left(g\right)}$ is convex in prices.
\end{assumption}
Utility maximization guarantees that the indirect utility function is quasi-convex and decreasing in prices. A convex function is quasi-convex, but not vice versa. \citet{Quah2000} shows that convex indirect utility functions can typically be observed for preferences defined over price-income combinations. From Assumption \ref{as2}, we obtain the sign restrictions
\begin{eqnarray*}
    \frac{\partial V^{\left(g\right)}}{\partial\tilde{p}_{\alpha,g}} & \leq & 0\\
    \frac{\partial^{2}V^{\left(g\right)}}{\partial\tilde{p}_{\alpha,g}^{2}} & \geq & 0.
\end{eqnarray*}

\begin{assumption}\label{as3}
    For all $g,$ $\left|\frac{\partial Q_{\alpha}^{\left(g\right)}}{\partial p_{\alpha}}\right|>\left|\frac{\partial Q_{\alpha}^{\left(g\right)}}{\partial p_{\beta}}\right|$ and $\left|\frac{\partial Q_{\beta}^{\left(g\right)}}{\partial p_{\beta}}\right|>\left|\frac{\partial Q_{\beta}^{\left(g\right)}}{\partial p_{\alpha}}\right|.$
\end{assumption}
This assumption requires that the demand responds more strongly to changes in its own price than to changes in cross-prices (\citet{Munro.et.al.2025}). From Assumption \ref{as3}, we can specify the sign of $\frac{\partial p_{\alpha}}{\partial W_{g}}.$

\begin{lem}\label{le3}
Suppose Assumption \ref{as3} holds. Then $\frac{\partial p_{\alpha}}{\partial W_{g}}>0$ for all $g.$
\end{lem}
Intuitively, when cross-price elasticities are sufficiently small relative to own-price elasticities, the market price of good $\alpha$
must increase as treatment is expanded. Under Assumptions \ref{as2} and \ref{as3}, the first term in Lemma \ref{le2} becomes
\begin{equation*}
    \underset{\left(+\right)}{\underbrace{\frac{\partial p_{\alpha}}{\partial W_{s}}}}\underset{\left(+\right)}{\underbrace{\frac{\partial p_{\alpha}}{\partial W_{r}}}}\sum_{g=1}^{G}\left(1-\tau_{\alpha}W_{g}\right)^{2}\underset{\left(+\right)}{\underbrace{\frac{\partial^{2}V^{\left(g\right)}}{\partial\tilde{p}_{\alpha,g}^{2}}}}>0.
\end{equation*}

\begin{assumption}\label{as4}
    For all $g,$ $\frac{-\tilde{p}_{\alpha,g}\frac{\partial^{2}V^{\left(g\right)}}{\partial\tilde{p}_{\alpha,g}^{2}}}{\frac{\partial V^{\left(g\right)}}{\partial\tilde{p}_{\alpha,g}}}\leq1.$
\end{assumption}
The expression on the left-hand side corresponds, in the context of consumer demand, to the Arrow-Pratt measure of relative risk aversion for the indirect utility function with respect to the coupon-adjusted price. Since $V^{\left(g\right)}$ is convex in prices, this measure is nonnegative. Assumption \ref{as4} restricts it to lie between 0 and 1, which ensures that the entire second component of Lemma \ref{le2},
\begin{equation*}
    -\tau_{\alpha}\left[\underset{\left(+\right)}{\underbrace{\frac{\partial p_{\alpha}}{\partial W_{s}}}}\underset{\left(+\right)}{\underbrace{\left(1-\frac{-\tilde{p}_{\alpha,r}\frac{\partial^{2}V^{\left(r\right)}}{\partial\tilde{p}_{\alpha,r}^{2}}}{\frac{\partial V^{\left(r\right)}}{\partial\tilde{p}_{\alpha,r}}}\right)}}\underset{\left(-\right)}{\underbrace{\frac{\partial V^{\left(r\right)}}{\partial\tilde{p}_{\alpha,r}}}}+\underset{\left(+\right)}{\underbrace{\frac{\partial p_{\alpha}}{\partial W_{r}}}}\underset{\left(+\right)}{\underbrace{\left(1-\frac{-\tilde{p}_{\alpha,s}\frac{\partial^{2}V^{\left(s\right)}}{\partial\tilde{p}_{\alpha,s}^{2}}}{\frac{\partial V^{\left(s\right)}}{\partial\tilde{p}_{\alpha,s}}}\right)}}\underset{\left(-\right)}{\underbrace{\frac{\partial V^{\left(s\right)}}{\partial\tilde{p}_{\alpha,s}}}}\right]
\end{equation*}
is positive.

In the Cobb-Douglas case, the Arrow-Pratt measure equals 1, which makes each parenthetical term equal to zero and causes the entire second component to vanish. Thus, Cobb-Douglas preferences form an important boundary case for the supermodularity of the welfare function.

\begin{assumption}\label{as5}
    For all distinct $s$ and r, $\frac{\partial^{2}p_{\alpha}}{\partial W_{s}\partial W_{r}}<0.$
\end{assumption}
This assumption requires that the equilibrium price of good $\alpha$ is submodular in the treatment vector. The third component in Lemma \ref{le2} becomes
\begin{equation*}
    \underset{\left(-\right)}{\underbrace{\frac{\partial^{2}p_{\alpha}}{\partial W_{s}\partial W_{r}}}}\sum_{g=1}^{G}\left[\underset{\left(-\right)}{\underbrace{\frac{\partial V^{\left(g\right)}}{\partial\tilde{p}_{\alpha,g}}}}\left(1-\tau_{\alpha}W_{g}\right)\right].
\end{equation*}

Because $\frac{\partial V^{\left(g\right)}}{\partial\tilde{p}_{\alpha,g}}\leq0$ and $1-\tau_{\alpha}W_{g}>0,$ the summation is nonpositive. Combined with $\frac{\partial^{2}p_{\alpha}}{\partial W_{s}\partial W_{r}}<0,$ the overall term is therefore positive.

In the Cobb-Douglas benchmark with the convex supply, Assumption \ref{as5} holds automatically. However, if both demand and supply are linear, the sign can reverse. To illustrate, consider
\begin{eqnarray*}
    Q_{\alpha}^{\left(g\right)} & = & \varphi_{g}-\gamma_{g}\left(1-\tau_{\alpha}W_{g}\right)p_{\alpha}\\
    S_{\alpha} & = & \delta_{\alpha}+\kappa_{\alpha}p_{\alpha}.
\end{eqnarray*}
In this linear case one can show that
\begin{equation*}
    \frac{\partial^{2}p_{\alpha}}{\partial W_{s}\partial W_{r}}>0
\end{equation*}
for all distinct $s$ and $r.$ Thus, the third component of Lemma \ref{le2} becomes negative rather than positive. The intuition is that when demand and supply are both linear, they are simultaneously convex and concave, leaving insufficient curvature to generate the negative cross-partial of equilibrium prices that is required for supermodularity. Sufficient convexity of demand and supply is therefore critical to preserve the desired complementarity structure.

\begin{thm}\label{th4}
Suppose that Assumptions \ref{as2}, \ref{as3}, \ref{as4}, and \ref{as5} hold. If we further assume $\frac{\partial Q_{\beta}^{\left(g\right)}}{\partial p_{\alpha}}=0$ for all $g,$ then the policymaker's value function $V\left(\boldsymbol{W}\right)$ is supermodular in $\boldsymbol{W}.$
\end{thm}

This chain of results shows that supermodularity is not confined to the Cobb-Douglas benchmark. It holds more generally under interpretable conditions: convex indirect utility, dominance of own-price responses, bounded relative risk aversion, and submodular equilibrium price responses. The Cobb-Douglas case serves as a boundary example where certain terms cancel exactly, making the complementarity structure especially transparent.

While Theorems \ref{th1} and \ref{th4} establish conditions under which the welfare function is supermodular, in practice the policymaker must rely on estimated primitives. Let $\theta$ denote the vector of structural parameters that govern demand and supply, and let $\hat{\theta}$ be an estimator constructed from the data. For the true parameter value $\theta^{*},$ let
\begin{equation*}
    \boldsymbol{W}^{*}\in\underset{\boldsymbol{W}\in\left\{ 0,1\right\} ^{G}}{\arg\max}V\left(\boldsymbol{W};\theta^{*}\right)
\end{equation*}
denote the true optimal allocation, and let
\begin{equation*}
    \hat{\boldsymbol{W}}\in\underset{\boldsymbol{W}\in\left\{ 0,1\right\} ^{G}}{\arg\max}V\left(\boldsymbol{W};\hat{\theta}\right)
\end{equation*}
denote the estimated optimal allocation. Define the risk as
\begin{equation*}
    \mathcal{R}\equiv V\left(\boldsymbol{W}^{*};\theta^{*}\right)-V\left(\hat{\boldsymbol{W}};\hat{\theta}\right)
\end{equation*}
With this notation in place, we now introduce additional assumptions to derive the expected welfare risk.

\begin{assumption}\label{as6}
    For each group's indirect utility function $V^{\left(g\right)},$ there exists a constant $\overline{S}_{g}<\infty$ such that for all $\boldsymbol{W}\in\left\{ 0,1\right\} ^{G}$ and for all $\theta,\theta^{'}\in\Theta,$
    \begin{equation*}
        \left|V^{\left(g\right)}\left(\boldsymbol{W};\theta^{'}\right)-V^{\left(g\right)}\left(\boldsymbol{W};\theta\right)\right|\leq\overline{S}_{g}\left\Vert \theta^{'}-\theta\right\Vert \qquad\mathrm{a}.\mathrm{s}.
    \end{equation*}
    and the parameter space $\Theta$ is compact.
\end{assumption}
This assumption ensures that the indirect utility functions are uniformly Lipschitz continuous in the parameter vector and that the parameter space is bounded, so that all admissible parameter values lie within a finite distance of one another.

\begin{assumption}\label{as7}
    There exists constants $c_{1},c_{2}>0$ such that, for all $u>0$ and sufficiently large sample size $n,$
    \begin{equation*}
        \mathrm{Pr}\left(\left\Vert \hat{\theta}-\theta^{*}\right\Vert >u\right)\leq c_{1}\exp\left(-c_{2}nu^{2}\right).
    \end{equation*}
\end{assumption}
This assumption specifies an exponential tail probability bound for the estimator. It requires that the probability of large estimation errors decays at an exponential rate in the sample size, which is satisfied by regular $\sqrt{n}$-consistent estimators such as the maximum likelihood estimator under standard regularity conditions.

\begin{thm}\label{th5}
Suppose that Assumptions \ref{as6} and \ref{as7} holds. For sufficiently large sample size $n$, the (pointwise) expected risk is written as
\begin{equation*}
    \mathbb{E}\left|\mathcal{R}\right|=O\left(n^{-\frac{1}{2}}\right).
\end{equation*}
\end{thm}

Theorem \ref{th5} establishes that the expected welfare loss from using an estimated policy instead of the true optimal policy decreases at the standard parametric rate of $n^{-\frac{1}{2}}.$ As the sample size increases, the welfare achieved by the plug-in allocation converges to the welfare under the true optimal allocation at this rate. This result indicates that, under Assumptions \ref{as6} and \ref{as7}, estimation uncertainty in the primitives translates proportionally into welfare uncertainty. The policymaker's expected welfare performance therefore improves at the same rate as the accuracy of the parameter estimates, showing that the plug-in allocation rule is asymptotically efficient in terms of welfare.

\section{Empirical Illustration}\label{s3}

This section implements the theoretical framework using household expenditure data. The aim is to calibrate the supply elasticity $\kappa_{\alpha}$ and group-level heterogeneity in income and preferences, which together determine the planner's welfare function. \citet{Zoutman.et.al.2018} provides theoretical evidence on how supply elasticity can be identified in settings where policy interventions only shift demand, since variation in an ad-valorem tax or its mirror image, a coupon subsidy, affects the pre-tax price. \citet{Munro.et.al.2025} implements this idea using simulated data, calibrated with moments from \citet{Filmer.et.al.2023}. We adapt their strategy to our setting.

Under Cobb-Douglas utility, the group-level demand for good $\alpha$ is
\begin{equation*}
    Q_{\alpha}^{\left(g\right)}=\frac{\gamma_{g}M_{g}}{\left(1-\tau_{\alpha}W_{g}\right)p_{\alpha}},
\end{equation*}
so that aggregate demand is
\begin{eqnarray*}
    Q_{\alpha} & = & \sum_{g=1}^{G}Q_{\alpha}^{\left(g\right)}\\
    & = & \frac{1}{p_{\alpha}}\sum_{g=1}^{G}\frac{\gamma_{g}M_{g}}{1-\tau_{\alpha}W_{g}}
\end{eqnarray*}
which implies
\begin{equation*}
    \log Q_{\alpha}^{\left(g\right)}=\log\left(\sum_{g=1}^{G}\frac{\gamma_{g}M_{g}}{1-\tau_{\alpha}W_{g}}\right)-\log p_{\alpha}.
\end{equation*}
On the supply side, we assume a constant-elasticity function,
\begin{equation*}
    \log S_{\alpha}=\log\delta_{\alpha}+\kappa_{\alpha}\log p_{\alpha},
\end{equation*}
so that equilibrium requires
\begin{equation*}
    \log\left(\sum_{g=1}^{G}\frac{\gamma_{g}M_{g}}{1-\tau_{\alpha}W_{g}}\right)-\log p_{\alpha}=\log\delta_{\alpha}+\kappa_{\alpha}\log p_{\alpha}.
\end{equation*}
Recall that in this benchmark the policymaker's welfare function takes the form
\begin{equation*}
    V\left(\boldsymbol{W}\right)=-\left\{ \sum_{g=1}^{G}\lambda_{g}\left[\gamma_{g}\log\left(1-\tau_{\alpha}W_{g}\right)\right]+\frac{\log\left(\sum_{g=1}^{G}\frac{\gamma_{g}M_{g}}{1-\tau_{\alpha}W_{g}}\right)}{1+\kappa_{\alpha}}\sum_{g=1}^{G}\lambda_{g}\gamma_{g}\right\}.
\end{equation*}
which shows that evaluating welfare requires knowledge of the supply elasticity $\kappa_{\alpha}.$

The data are drawn from \citet{Filmer.et.al.2023} and contain household expenditure information from the Philippines. The analysis focuses on eggs as the representative good $\alpha,$ for which both prices and expenditures are observed across treated and untreated households. Income is measured using the logged proxy means-test score, and the food expenditure share is used to capture the budget composition.

An important advantage of this dataset is that the supply elasticity of eggs can be identified. The coupon subsidy shifts demand along the existing supply curve without directly affecting the supply curve itself. This allows estimation of the supply elasticity. \citet{Munro.et.al.2025} reports an implied elasticity of $\kappa_{\alpha}=1.11.$

The planner's welfare function also depends on group-specific income $M_{g}$ and expenditure share $\gamma_{g},$ which we calibrate using the heterogeneity documented in \citet{Filmer.et.al.2023}. Together, these calibrated parameters allow us to evaluate welfare under different allocation rules and to illustrate how the theoretical framework can be applied empirically.

\begin{figure}[htbp]
\begin{centering}
\includegraphics{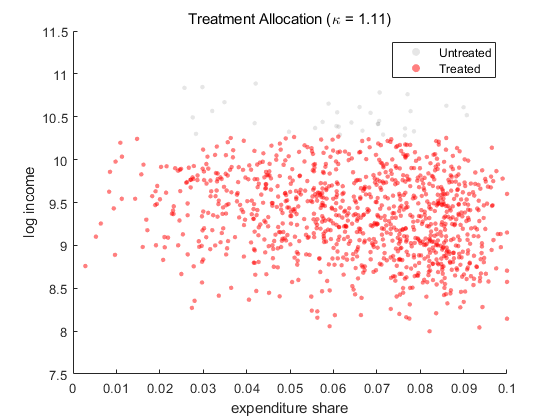}
\par\end{centering}
\caption{Treatment Allocations when $\kappa_{\alpha}=1.11$}
\label{fig1}
\end{figure}

\begin{figure}[htbp]
\begin{centering}
\includegraphics{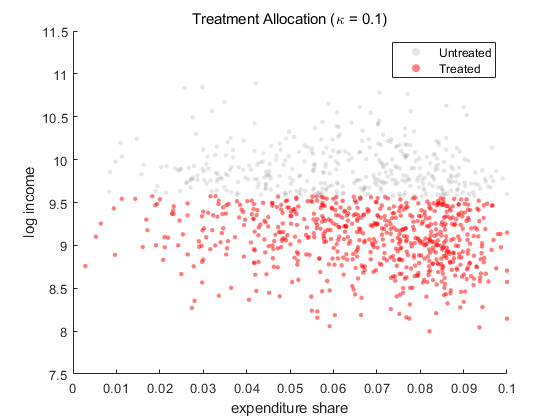}
\par\end{centering}
\caption{Treatment Allocations when $\kappa_{\alpha}=0.1$}
\label{fig2}
\end{figure}

\begin{figure}[htbp]
\begin{centering}
\includegraphics{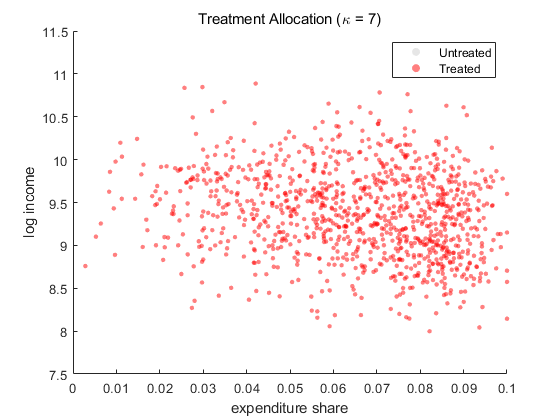}
\par\end{centering}
\caption{Treatment Allocations when $\kappa_{\alpha}=7$}
\label{fig3}
\end{figure}

The sample includes 1,058 households that report both income and the expenditure share on food items. In our empirical setting, the planner is assumed to consider a 10 percent discount coupon for the good of interest. Given the heterogeneity in demand across individuals from the data, an optimal allocation is implemented. Figures \ref{fig1}-\ref{fig3} illustrate the optimal treatment allocations under different values of the supply elasticity parameter $\kappa_{\alpha}.$

When the supply elasticity is moderate ($\kappa_{\alpha}=1.11$), the planner treats a large fraction of households. In this case, price adjustments are present but not severe, so expanding treatment generates welfare gains for many groups without imposing substantial negative spillovers on the untreated.

When the supply elasticity is low ($\kappa_{\alpha}=0.1$), indicating an inelastic supply, corresponding to an inelastic supply, the optimal allocation becomes much more selective. The planner primarily targets lower-income households. The intuition is that a broad distribution of coupons would significantly increase equilibrium prices, thereby eroding welfare for untreated households and partially offsetting the benefits for treated ones. As a result, targeting is used to concentrate gains where marginal utility is highest while limiting adverse price effects.

In contrast, when the supply elasticity is high ($\kappa_{\alpha}=7$), the planner finds it optimal to distribute coupons universally. With highly elastic supply, increases in demand are absorbed mainly through higher quantities rather than higher prices. Consequently, the general equilibrium price distortion is minimal, and the planner can expand treatment broadly without generating meaningful negative spillovers.

Finally, the allocation pattern appears to depend almost entirely on income. This arises because the expenditure share on eggs exhibits limited variation across households. Even higher-income households do not allocate a substantially larger fraction of their budget to eggs, reflecting the fact that eggs are a necessity with relatively stable budget shares. As a result, heterogeneity in welfare gains is driven primarily by income differences rather than variation in expenditure shares, leading the planner's allocation rule to effectively sort households by income.

We compare several alternative treatment assignment rules relative to the optimal policy, which serves as the benchmark. In particular, we consider three alternative rules: (\mbox{i}) no treatment, (\mbox{ii}) universal treatment, and (\mbox{iii}) a complement rule that assigns treatment only to households that are not treated under the optimal allocation. Table \ref{tab1} summarizes the results.

\begin{table}[htbp]
\caption{Various Treatment Allocations}
\label{tab1}
\centering
\begin{tabular}{@{}ccccc@{}}
\hline
& No-one & Optimal & Everyone & Complement \\
\hline
Welfare Gain       & 0 & 3.897 & 3.818 & $-$0.110 \\
Treated Prop. (\%) & 0 & 96.7  & 100   & 3.3        \\
\hline
\end{tabular}
\end{table}

Relative to the optimal policy, all alternative rules perform worse in terms of welfare. Treating everyone still yields higher welfare than treating no one, indicating that the intervention is beneficial on average. However, it falls short of the optimal allocation, which carefully balances the benefits of treatment against the general equilibrium price effects. In contrast, the complement rule performs particularly poorly, generating even lower welfare than the no-treatment benchmark. This highlights an important policy lesson: assigning treatment to inappropriate groups can be more harmful than not intervening at all. In this setting, misallocation exacerbates price distortions without delivering sufficient benefits to compensate for them.

As shown in the Figures \ref{fig1}-\ref{fig3}, the proportion of the treated household
varies as the supply elasticities changes. Table \ref{tab2} presents the corresponding
optimal treatment shares across different values of supply elasticity.

When supply is highly inelastic, the optimal policy treats a relatively
small fraction of households. This reflects the planner's concern
about price increases: expanding treatment raises demand, which, under
inelastic supply, translates primarily into higher prices rather than
higher quantities. These price increases impose negative spillovers
on untreated households and dampen the overall welfare gains.

As supply becomes more elastic, the optimal treatment share rises
steadily. With greater supply responsiveness, increases in demand
are absorbed more through quantities and less through prices, reducing
the magnitude of equilibrium distortions. This allows the planner
to extend treatment more broadly without generating substantial adverse
effects. When supply is sufficiently elastic, universal treatment
becomes optimal, as price distortions are minimal and the gains from
expanding coverage dominate.

Overall, these results show that the effectiveness and desirability
of different treatment rules depend crucially on supply conditions.
Accounting for general equilibrium effects is therefore essential
for designing welfare-maximizing policies.

\begin{table}[htbp]
\caption{Optimal Treatment Allocation with Various Supply Elasticities}
\label{tab2}
\centering
\begin{tabular}{@{}cccccccc@{}}
\hline
Sup. Elas.          & 0.05 & 0.2  & 0.5  & 1    & 1.11 & 3    & 5   \\
\hline
Treated Prop. (\%)  & 60.9 & 70.6 & 85.6 & 95.5 & 96.7 & 99.8 & 100 \\
\hline
\end{tabular}
\end{table}

\section{Conclusion}\label{s4}

This paper studies how to allocate subsidies when interventions shift market prices and affect both treated and untreated groups. We show that the planner's welfare function is supermodular under broad conditions, which makes the allocation problem tractable and delivers clear comparative statics. In the Cobb-Douglas benchmark, closed-form welfare expressions highlight how supply elasticity and income distribution determine whether universal or targeted subsidies are optimal. We establish statistical guarantees that account for estimation uncertainty and provide an empirical illustration that calibrates key parameters using household expenditure data. While our analysis is motivated by subsidy allocation, the framework applies more broadly to policy interventions that affect market-clearing prices and thereby generate spillovers across treated and untreated groups. Together, these results demonstrate how incorporating equilibrium price responses yields both tractable analysis and clear guidance for policy design.

%%%%%%%%%%%%%%%%%%%%%%%%%%%%%%%%%%%%%%%%%%%%%%
%% Example with single Appendix:            %%
%%%%%%%%%%%%%%%%%%%%%%%%%%%%%%%%%%%%%%%%%%%%%%
\appendix
\section{Proofs of Results}\label{appn}
\subsection{Proof of Theorem \ref{th1}}
The demand for group $g$ is derived as
\begin{eqnarray*}
    Q_{\alpha}^{\left(g\right)} & = & \frac{\gamma_{g}M_{g}}{\left(1-\tau_{\alpha}W_{g}\right)p_{\alpha}}\\
    Q_{\beta}^{\left(g\right)} & = & \frac{\left(1-\gamma_{g}\right)M_{g}}{p_{\beta}}.
\end{eqnarray*}
The market clearing condition is given as
\begin{equation*}
    Q\equiv\sum_{g=1}^{G}Q^{\left(g\right)}=S.
\end{equation*}
The aggregate demand $Q$ is expressed as
\begin{eqnarray*}
    Q & = & \left[\begin{array}{c}
\sum_{g=1}^{G}Q_{\alpha}^{\left(g\right)}\\
\sum_{g=1}^{G}Q_{\beta}^{\left(g\right)}
\end{array}\right]\\
    & = & \left[\begin{array}{c}
\sum_{g=1}^{G}\frac{\gamma_{g}M_{g}}{\left(1-\tau_{\alpha}W_{g}\right)p_{\alpha}}\\
\sum_{g=1}^{G}\frac{\left(1-\gamma_{g}\right)M_{g}}{p_{\beta}}
\end{array}\right]\\
    & = & \left[\begin{array}{c}
\frac{1}{p_{\alpha}}\sum_{g=1}^{G}\frac{\gamma_{g}M_{g}}{\left(1-\tau_{\alpha}W_{g}\right)}\\
\frac{1}{p_{\beta}}\sum_{g=1}^{G}\left(1-\gamma_{g}\right)M_{g}
\end{array}\right].
\end{eqnarray*}
By taking log, we observe
\begin{equation*}
    \left[\begin{array}{c}
\log Q_{\alpha}\\
\log Q_{\beta}
\end{array}\right]=\left[\begin{array}{c}
-\log p_{\alpha}+\log\left(\sum_{g=1}^{G}\frac{\gamma_{g}M_{g}}{1-\tau_{\alpha}W_{g}}\right)\\
-\log p_{\beta}+\log\left(\sum_{g=1}^{G}\left(1-\gamma_{g}\right)M_{g}\right)
\end{array}\right].
\end{equation*}
Given the supply functions
\begin{equation*}
    \left[\begin{array}{c}
\log S_{\alpha}\\
\log S_{\beta}
\end{array}\right]=\left[\begin{array}{c}
\log\delta_{\alpha}+\kappa_{\alpha}\log p_{\alpha}\\
\log\delta_{\beta}+\kappa_{\beta}\log p_{\beta}
\end{array}\right],
\end{equation*}
the market price is determined as
\begin{eqnarray*}
    \left(1+\kappa_{\alpha}\right)\log p_{\alpha} & = & \log\left(\sum_{g=1}^{G}\frac{\gamma_{g}M_{g}}{1-\tau_{\alpha}W_{g}}\right)-\log\delta_{\alpha}\\
    \left(1+\kappa_{\beta}\right)\log p_{\beta} & = & \log\left(\sum_{g=1}^{G}\left(1-\gamma_{g}\right)M_{g}\right)-\log\delta_{\beta}
\end{eqnarray*}
so that
\begin{eqnarray*}
    p^{*} & = & \left[\begin{array}{c}
\exp\left(\frac{\log\left(\frac{1}{\delta_{\alpha}}\sum_{g=1}^{G}\frac{\gamma_{g}M_{g}}{1-\tau_{\alpha}W_{g}}\right)}{1+\kappa_{\alpha}}\right)\\
\exp\left(\frac{\log\left(\frac{1}{\delta_{\beta}}\sum_{g=1}^{G}\left(1-\gamma_{g}\right)M_{g}\right)}{1+\kappa_{\beta}}\right)
\end{array}\right]\\
    & = & \left[\begin{array}{c}
\left(\frac{1}{\delta_{\alpha}}\sum_{g=1}^{G}\frac{\gamma_{g}M_{g}}{1-\tau_{\alpha}W_{g}}\right)^{\frac{1}{1+\kappa_{\alpha}}}\\
\left(\frac{1}{\delta_{\beta}}\sum_{g=1}^{G}\left(1-\gamma_{g}\right)M_{g}\right)^{\frac{1}{1+\kappa_{\beta}}}
\end{array}\right].
\end{eqnarray*}
Then, by plugging $p^{*}$ and $Q^{\left(g\right)}\left(p^{*}\left(\boldsymbol{W}\right)\right)$ into the value function, we obtain
\begin{eqnarray*}
    V_{g}\left(\boldsymbol{W}\right) & = & \gamma_{g}\log Q_{\alpha}^{\left(g\right)}+\left(1-\gamma_{g}\right)\log\left(Q_{\beta}^{\left(g\right)}\right)\\
    & = & \gamma_{g}\log\frac{\gamma_{g}M_{g}}{\left(1-\tau_{\alpha}W_{g}\right)p_{\alpha}\left(\boldsymbol{W}\right)}+\left(1-\gamma_{g}\right)\log\left(\frac{\left(1-\gamma_{g}\right)M_{g}}{p_{\beta}}\right)\\
    & = & \gamma_{g}\left[\log\gamma_{g}M_{g}-\log\left(1-\tau_{\alpha}W_{g}\right)-\log p_{\alpha}\left(\boldsymbol{W}\right)\right]\\
    &  & +\left(1-\gamma_{g}\right)\left[\log\left(1-\gamma_{g}\right)M_{g}-\log p_{\beta}\right].
\end{eqnarray*}
Then, the social planner maximizes the sum of individual indirect utilities
\begin{eqnarray*}
    V\left(\boldsymbol{W}\right) & = & \sum_{g=1}^{G}\lambda_{g}V_{g}\left(\boldsymbol{W}\right)\\
    & = & \sum_{g=1}^{G}\lambda_{g}\left\{ \gamma_{g}\left[\log\gamma_{g}M_{g}-\log\left(1-\tau_{\alpha}W_{g}\right)-\log p_{\alpha}\left(\boldsymbol{W}\right)\right]\right\} \\
    &  & +\sum_{g=1}^{G}\lambda_{g}\left\{ \left(1-\gamma_{g}\right)\left[\log\left(1-\gamma_{g}\right)M_{g}-\log p_{\beta}\right]\right\} \\
    & = & \sum_{g=1}^{G}\lambda_{g}\gamma_{g}\left\{ -\log\left(1-\tau_{\alpha}W_{g}\right)-\log p_{\alpha}\left(\boldsymbol{W}\right)\right\} \\
    &  & +\sum_{g=1}^{G}\lambda_{g}\left[\gamma_{g}\log\gamma_{g}M_{g}+\left(1-\gamma_{g}\right)\left[\log\left(1-\gamma_{g}\right)M_{g}-\log p_{\beta}\right]\right].
\end{eqnarray*}
The last line does not depend on $\boldsymbol{W},$ so dropping it does not change the maximizer. Then,
\begin{eqnarray*}
    V\left(\boldsymbol{W}\right) & = & \sum_{g=1}^{G}\lambda_{g}\gamma_{g}\left\{ -\log\left(1-\tau_{\alpha}W_{g}\right)-\log p_{\alpha}\left(\boldsymbol{W}\right)\right\} \\
    & = & -\sum_{g=1}^{G}\lambda_{g}\left[\gamma_{g}\log\left(1-\tau_{\alpha}W_{g}\right)\right]-\log p_{\alpha}\left(\boldsymbol{W}\right)\sum_{g=1}^{G}\lambda_{g}\gamma_{g}\\
    & = & -\sum_{g=1}^{G}\lambda_{g}\left[\gamma_{g}\log\left(1-\tau_{\alpha}W_{g}\right)\right]-\frac{\log\left(\frac{1}{\delta_{\alpha}}\sum_{g=1}^{G}\frac{\gamma_{g}M_{g}}{1-\tau_{\alpha}W_{g}}\right)}{1+\kappa_{\alpha}}\sum_{g=1}^{G}\lambda_{g}\gamma_{g}\\
    & = & -\left\{ \sum_{g=1}^{G}\lambda_{g}\left[\gamma_{g}\log\left(1-\tau_{\alpha}W_{g}\right)\right]+\frac{\log\left(\frac{1}{\delta_{\alpha}}\sum_{g=1}^{G}\frac{\gamma_{g}M_{g}}{1-\tau_{\alpha}W_{g}}\right)}{1+\kappa_{\alpha}}\sum_{g=1}^{G}\lambda_{g}\gamma_{g}\right\} \\
    & = & -\left\{ \sum_{g=1}^{G}\lambda_{g}\left[\gamma_{g}\log\left(1-\tau_{\alpha}W_{g}\right)\right]+\frac{\log\left(\sum_{g=1}^{G}\frac{\gamma_{g}M_{g}}{1-\tau_{\alpha}W_{g}}\right)-\log\left(\delta_{\alpha}\right)}{1+\kappa_{\alpha}}\sum_{g=1}^{G}\lambda_{g}\gamma_{g}\right\} \\
    & = & -\left\{ \sum_{g=1}^{G}\lambda_{g}\left[\gamma_{g}\log\left(1-\tau_{\alpha}W_{g}\right)\right]+\frac{\log\left(\sum_{g=1}^{G}\frac{\gamma_{g}M_{g}}{1-\tau_{\alpha}W_{g}}\right)}{1+\kappa_{\alpha}}\sum_{g=1}^{G}\lambda_{g}\gamma_{g}\right\} \\
    &  & +\frac{\log\left(\delta_{\alpha}\right)}{1+\kappa_{\alpha}}\sum_{g=1}^{G}\lambda_{g}\gamma_{g}
\end{eqnarray*}
where the last term does not depend on $\boldsymbol{W}.$ Thus,
\begin{equation*}
    V\left(\boldsymbol{W}\right)=-\left\{ \sum_{g=1}^{G}\lambda_{g}\left[\gamma_{g}\log\left(1-\tau_{\alpha}W_{g}\right)\right]+\frac{\log\left(\sum_{g=1}^{G}\frac{\gamma_{g}M_{g}}{1-\tau_{\alpha}W_{g}}\right)}{1+\kappa_{\alpha}}\sum_{g=1}^{G}\lambda_{g}\gamma_{g}\right\} .
\end{equation*}
The first order derivative w.r.t $W_{r}$ for any group $r$ is
\begin{eqnarray*}
    \frac{\partial V}{\partial W_{r}} & = & -\left\{ \frac{-\tau_{\alpha}\lambda_{r}\gamma_{r}}{1-\tau_{\alpha}W_{r}}+\frac{\sum_{g=1}^{G}\lambda_{g}\gamma_{g}}{1+\kappa_{\alpha}}\frac{1}{\sum_{g=1}^{G}\frac{\gamma_{g}M_{g}}{1-\tau_{\alpha}W_{g}}}\frac{-\gamma_{r}M_{r}*\left(-\tau_{\alpha}\right)}{\left(1-\tau_{\alpha}W_{r}\right)^{2}}\right\} \\
    & = & \frac{\tau_{\alpha}\lambda_{r}\gamma_{r}}{1-\tau_{\alpha}W_{r}}-\frac{\sum_{g=1}^{G}\lambda_{g}\gamma_{g}}{1+\kappa_{\alpha}}\frac{1}{\sum_{g=1}^{G}\frac{\gamma_{g}M_{g}}{1-\tau_{\alpha}W_{g}}}\frac{\gamma_{r}M_{r}\tau_{\alpha}}{\left(1-\tau_{\alpha}W_{r}\right)^{2}}\\
    & = & \frac{\tau_{\alpha}\gamma_{r}}{\left(1-\tau_{\alpha}W_{r}\right)^{2}}\left[\lambda_{r}\left(1-\tau_{\alpha}W_{r}\right)-\frac{M_{r}}{1+\kappa_{\alpha}}\sum_{g=1}^{G}\lambda_{g}\gamma_{g}\frac{1}{\sum_{g=1}^{G}\frac{\gamma_{g}M_{g}}{1-\tau_{\alpha}W_{g}}}\right].
\end{eqnarray*}
The second order derivative w.r.t $W_{r}$ and $W_{s}$ for any groups $r\neq s,$ is
\begin{eqnarray*}
    &  & \frac{\partial^{2}V}{\partial W_{s}\partial W_{r}}\\
    & = & \frac{\partial}{\partial W_{s}}\left[\frac{\tau_{\alpha}\gamma_{r}}{\left(1-\tau_{\alpha}W_{r}\right)^{2}}\left\{ \lambda_{r}\left(1-\tau_{\alpha}W_{r}\right)-\frac{M_{r}}{1+\kappa_{\alpha}}\left(\sum_{g=1}^{G}\lambda_{g}\gamma_{g}\right)\frac{1}{\sum_{g=1}^{G}\frac{\gamma_{g}M_{g}}{1-\tau_{\alpha}W_{g}}}\right\} \right]\\
    & = & -\frac{\tau_{\alpha}\gamma_{r}}{\left(1-\tau_{\alpha}W_{r}\right)^{2}}\frac{M_{r}}{1+\kappa_{\alpha}}\left(\sum_{g=1}^{G}\lambda_{g}\gamma_{g}\right)\frac{-\frac{-\gamma_{s}M_{s}*\left(-\tau_{\alpha}\right)}{\left(1-\tau_{\alpha}W_{s}\right)^{2}}}{\left(\sum_{g=1}^{G}\frac{\gamma_{g}M_{g}}{1-\tau_{\alpha}W_{g}}\right)^{2}}\\
    & = & \frac{\tau_{\alpha}\gamma_{r}}{\left(1-\tau_{\alpha}W_{r}\right)^{2}}\frac{\gamma_{s}M_{s}\tau_{\alpha}}{\left(1-\tau_{\alpha}W_{s}\right)^{2}}\frac{M_{r}}{1+\kappa_{\alpha}}\frac{\sum_{g=1}^{G}\lambda_{g}\gamma_{g}}{\left(\sum_{g=1}^{G}\frac{\gamma_{g}M_{g}}{1-\tau_{\alpha}W_{g}}\right)^{2}}\\
    & = & \frac{\tau_{\alpha}^{2}\gamma_{r}\gamma_{s}M_{r}M_{s}}{\left(1+\kappa_{\alpha}\right)\left(1-\tau_{\alpha}W_{r}\right)^{2}\left(1-\tau_{\alpha}W_{s}\right)^{2}}\frac{\sum_{g=1}^{G}\lambda_{g}\gamma_{g}}{\left(\sum_{g=1}^{G}\frac{\gamma_{g}M_{g}}{1-\tau_{\alpha}W_{g}}\right)^{2}}\\
    & > & 0
\end{eqnarray*}
which concludes the supermodularity of $V\left(\boldsymbol{W}\right).$

\subsection{Proof of Lemma \ref{le2}}
Recall that
\begin{eqnarray*}
    &  & V\left(\boldsymbol{W}\right)\\
    & = & \sum_{g=1}^{G}V^{\left(g\right)}\left(\boldsymbol{W}\right)\\
    & = & \sum_{g=1}^{G}U^{\left(g\right)}\left(Q_{\alpha}^{\left(g\right)}\left(\left(1-\tau_{\alpha}W_{g}\right)p_{\alpha}\left(\boldsymbol{W}\right),p_{\beta}\left(\boldsymbol{W}\right)\right),Q_{\beta}^{\left(g\right)}\left(\left(1-\tau_{\alpha}W_{g}\right)p_{\alpha}\left(\boldsymbol{W}\right),p_{\beta}\left(\boldsymbol{W}\right)\right)\right)\\
    & = & \sum_{g=1}^{G}U^{\left(g\right)}\left(Q_{\alpha}^{\left(g\right)}\left(\tilde{p}_{\alpha,g}\left(\boldsymbol{W}\right),p_{\beta}\left(\boldsymbol{W}\right)\right),Q_{\beta}^{\left(g\right)}\left(\tilde{p}_{\alpha,g}\left(\boldsymbol{W}\right),p_{\beta}\left(\boldsymbol{W}\right)\right)\right).
\end{eqnarray*}
For any group $r,$ the first-order derivative is
\begin{equation*}
    \frac{\partial}{\partial W_{r}}\sum_{g=1}^{G}V^{\left(g\right)}\left(\boldsymbol{W}\right)=\sum_{g=1}^{G}\left[\frac{\partial U^{\left(g\right)}}{\partial\tilde{p}_{\alpha,g}}\frac{\partial\tilde{p}_{\alpha,g}}{\partial W_{r}}+\frac{\partial U^{\left(g\right)}}{\partial p_{\beta}}\frac{\partial p_{\beta}}{\partial W_{r}}\right],
\end{equation*}
where $\tilde{p}_{\alpha,g}=\left(1-\tau_{\alpha}W_{g}\right)p_{\alpha}.$ This can be written as
\begin{eqnarray*}
    \frac{\partial}{\partial W_{r}}\sum_{g=1}^{G}V^{\left(g\right)}\left(\boldsymbol{W}\right) & = & \sum_{g=1}^{G}\left[\frac{\partial U^{\left(g\right)}}{\partial\tilde{p}_{\alpha,g}}\left(\left(1-\tau_{\alpha}W_{g}\right)\frac{\partial p_{\alpha}}{\partial W_{r}}-\tau_{\alpha}p_{\alpha}\frac{\partial W_{g}}{\partial W_{r}}\right)+\frac{\partial U^{\left(g\right)}}{\partial p_{\beta}}\frac{\partial p_{\beta}}{\partial W_{r}}\right]\\
    & = & -\tau_{\alpha}p_{\alpha}\frac{\partial U^{\left(r\right)}}{\partial\tilde{p}_{\alpha,r}}+\frac{\partial p_{\alpha}}{\partial W_{r}}\sum_{g=1}^{G}\left[\frac{\partial U^{\left(g\right)}}{\partial\tilde{p}_{\alpha,g}}\left(1-\tau_{\alpha}W_{g}\right)\right]+\frac{\partial p_{\beta}}{\partial W_{r}}\sum_{g=1}^{G}\frac{\partial U^{\left(g\right)}}{\partial p_{\beta}}.
\end{eqnarray*}
Define
\begin{eqnarray*}
    J_{s}^{\left(g\right)} & \equiv & \left[\begin{array}{c}
\frac{\partial\tilde{p}_{\alpha,g}}{\partial W_{s}}\\
\frac{\partial p_{\beta}}{\partial W_{s}}
\end{array}\right]\\
    J_{r}^{\left(g\right)} & \equiv & \left[\begin{array}{c}
\frac{\partial\tilde{p}_{\alpha,g}}{\partial W_{r}}\\
\frac{\partial p_{\beta}}{\partial W_{r}}
\end{array}\right]\\
    H^{\left(g\right)} & \equiv & \left[\begin{array}{cc}
\frac{\partial^{2}V^{\left(g\right)}}{\partial\tilde{p}_{\alpha,g}^{2}} & \frac{\partial^{2}V^{\left(g\right)}}{\partial p_{\beta}\partial\tilde{p}_{\alpha,g}}\\
\frac{\partial^{2}V^{\left(g\right)}}{\partial\tilde{p}_{\alpha,g}\partial p_{\beta}} & \frac{\partial^{2}V^{\left(g\right)}}{\partial p_{\beta}^{2}}
\end{array}\right].
\end{eqnarray*}
The second-order mixed derivative is then
\begin{eqnarray*}
    &  & \frac{\partial^{2}}{\partial W_{s}\partial W_{r}}\sum_{g=1}^{G}V^{\left(g\right)}\left(\boldsymbol{W}\right)\\
    & = & \frac{\partial}{\partial W_{s}}\left(-\tau_{\alpha}p_{\alpha}\frac{\partial U^{\left(r\right)}}{\partial\tilde{p}_{\alpha,r}}+\frac{\partial p_{\alpha}}{\partial W_{r}}\sum_{g=1}^{G}\left[\frac{\partial U^{\left(g\right)}}{\partial\tilde{p}_{\alpha,g}}\left(1-\tau_{\alpha}W_{g}\right)\right]+\frac{\partial p_{\beta}}{\partial W_{r}}\sum_{g=1}^{G}\frac{\partial U^{\left(g\right)}}{\partial p_{\beta}}\right)\\
    & = & -\tau_{\alpha}\frac{\partial p_{\alpha}}{\partial W_{s}}\frac{\partial U^{\left(r\right)}}{\partial\tilde{p}_{\alpha,r}}-\tau_{\alpha}p_{\alpha}\left(\frac{\partial^{2}U^{\left(r\right)}}{\partial\tilde{p}_{\alpha,r}^{2}}\frac{\partial\tilde{p}_{\alpha,r}}{\partial W_{s}}+\frac{\partial^{2}U^{\left(r\right)}}{\partial p_{\beta}\partial\tilde{p}_{\alpha,r}}\frac{\partial p_{\beta}}{\partial W_{s}}\right)\\
    &  & +\frac{\partial^{2}p_{\alpha}}{\partial W_{s}\partial W_{r}}\sum_{g=1}^{G}\left[\frac{\partial U^{\left(g\right)}}{\partial\tilde{p}_{\alpha,g}}\left(1-\tau_{\alpha}W_{g}\right)\right]\\
    &  & +\frac{\partial p_{\alpha}}{\partial W_{r}}\left(\sum_{g=1}^{G}\left[\left(\frac{\partial^{2}U^{\left(g\right)}}{\partial\tilde{p}_{\alpha,g}^{2}}\frac{\partial\tilde{p}_{\alpha,g}}{\partial W_{s}}+\frac{\partial^{2}U^{\left(g\right)}}{\partial p_{\beta}\partial\tilde{p}_{\alpha,g}}\frac{\partial p_{\beta}}{\partial W_{s}}\right)\left(1-\tau_{\alpha}W_{g}\right)\right]-\frac{\partial U^{\left(s\right)}}{\partial\tilde{p}_{\alpha,s}}\tau_{\alpha}\right)\\
    &  & +\frac{\partial^{2}p_{\beta}}{\partial W_{s}\partial W_{r}}\sum_{g=1}^{G}\frac{\partial U^{\left(g\right)}}{\partial p_{\beta}}\\
    &  & +\frac{\partial p_{\beta}}{\partial W_{r}}\sum_{g=1}^{G}\left(\frac{\partial^{2}U^{\left(g\right)}}{\partial p_{\beta}^{2}}\frac{\partial p_{\beta}}{\partial W_{s}}+\frac{\partial^{2}U^{\left(g\right)}}{\partial\tilde{p}_{\alpha,g}\partial p_{\beta}}\frac{\partial\tilde{p}_{\alpha,g}}{\partial W_{s}}\right)\\
    & = & \sum_{g=1}^{G}J_{s}^{\left(g\right)T}H^{\left(g\right)}J_{r}^{\left(g\right)}\\
    &  & +\frac{\partial^{2}p_{\alpha}}{\partial W_{s}\partial W_{r}}\sum_{g=1}^{G}\left[\frac{\partial U^{\left(g\right)}}{\partial\tilde{p}_{\alpha,g}}\left(1-\tau_{\alpha}W_{g}\right)\right]\\
    &  & +\frac{\partial^{2}p_{\beta}}{\partial W_{s}\partial W_{r}}\sum_{g=1}^{G}\frac{\partial U^{\left(g\right)}}{\partial p_{\beta}}\\
    &  & -\tau_{\alpha}\left(\frac{\partial p_{\alpha}}{\partial W_{s}}\frac{\partial U^{\left(r\right)}}{\partial\tilde{p}_{\alpha,r}}+\frac{\partial p_{\alpha}}{\partial W_{r}}\frac{\partial U^{\left(s\right)}}{\partial\tilde{p}_{\alpha,s}}\right)
\end{eqnarray*}
where we used
\begin{eqnarray*}
    &  & \frac{\partial p_{\alpha}}{\partial W_{r}}\left(\sum_{g=1}^{G}\left[\left(\frac{\partial^{2}U^{\left(g\right)}}{\partial\tilde{p}_{\alpha,g}^{2}}\frac{\partial\tilde{p}_{\alpha,g}}{\partial W_{s}}+\frac{\partial^{2}U^{\left(g\right)}}{\partial p_{\beta}\partial\tilde{p}_{\alpha,g}}\frac{\partial p_{\beta}}{\partial W_{s}}\right)\left(1-\tau_{\alpha}W_{g}\right)\right]\right)\\
    & = & \left(\sum_{g=1}^{G}\left[\left(\frac{\partial^{2}U^{\left(g\right)}}{\partial\tilde{p}_{\alpha,g}^{2}}\frac{\partial\tilde{p}_{\alpha,g}}{\partial W_{s}}+\frac{\partial^{2}U^{\left(g\right)}}{\partial p_{\beta}\partial\tilde{p}_{\alpha,g}}\frac{\partial p_{\beta}}{\partial W_{s}}\right)\frac{\partial p_{\alpha}}{\partial W_{r}}\left(1-\tau_{\alpha}W_{g}\right)\right]\right)\\
    & = & \left(\sum_{g=1}^{G}\left[\left(\frac{\partial^{2}U^{\left(g\right)}}{\partial\tilde{p}_{\alpha,g}^{2}}\frac{\partial\tilde{p}_{\alpha,g}}{\partial W_{s}}+\frac{\partial^{2}U^{\left(g\right)}}{\partial p_{\beta}\partial\tilde{p}_{\alpha,g}}\frac{\partial p_{\beta}}{\partial W_{s}}\right)\left(\tau_{\alpha}p_{\alpha}\frac{\partial W_{g}}{\partial W_{r}}+\frac{\partial\tilde{p}_{\alpha,g}}{\partial W_{r}}\right)\right]\right)\\
    & = & \sum_{g=1}^{G}\frac{\partial\tilde{p}_{\alpha,g}}{\partial W_{r}}\left(\frac{\partial^{2}U^{\left(g\right)}}{\partial\tilde{p}_{\alpha,g}^{2}}\frac{\partial\tilde{p}_{\alpha,g}}{\partial W_{s}}+\frac{\partial^{2}U^{\left(g\right)}}{\partial p_{\beta}\partial\tilde{p}_{\alpha,g}}\frac{\partial p_{\beta}}{\partial W_{s}}\right)\\
    &  & +\tau_{\alpha}p_{\alpha}\left(\frac{\partial^{2}U^{\left(r\right)}}{\partial\tilde{p}_{\alpha,r}^{2}}\frac{\partial\tilde{p}_{\alpha,r}}{\partial W_{s}}+\frac{\partial^{2}U^{\left(r\right)}}{\partial p_{\beta}\partial\tilde{p}_{\alpha,r}}\frac{\partial p_{\beta}}{\partial W_{s}}\right).
\end{eqnarray*}
This establishes the first part of Lemma \ref{le2}.

Moreover, if $\frac{\partial Q_{\beta}^{\left(g\right)}}{\partial p_{\alpha}}=0$ for all $g,$ then
\begin{eqnarray*}
    \frac{\partial^{2}}{\partial W_{s}\partial W_{r}}\sum_{g=1}^{G}V^{\left(g\right)}\left(\boldsymbol{W}\right) & = & \sum_{g=1}^{G}J_{s}^{\left(g\right)T}H^{\left(g\right)}J_{r}^{\left(g\right)}\\
    &  & +\frac{\partial^{2}p_{\alpha}}{\partial W_{s}\partial W_{r}}\sum_{g=1}^{G}\left[\frac{\partial U^{\left(g\right)}}{\partial\tilde{p}_{\alpha,g}}\left(1-\tau_{\alpha}W_{g}\right)\right]\\
    &  & +\frac{\partial^{2}p_{\beta}}{\partial W_{s}\partial W_{r}}\sum_{g=1}^{G}\frac{\partial U^{\left(g\right)}}{\partial p_{\beta}}\\
    &  & -\tau_{\alpha}\left(\frac{\partial p_{\alpha}}{\partial W_{s}}\frac{\partial U^{\left(r\right)}}{\partial\tilde{p}_{\alpha,r}}+\frac{\partial p_{\alpha}}{\partial W_{r}}\frac{\partial U^{\left(s\right)}}{\partial\tilde{p}_{\alpha,s}}\right)\\
    & = & \sum_{g=1}^{G}\frac{\partial\tilde{p}_{\alpha,g}}{\partial W_{s}}\frac{\partial\tilde{p}_{\alpha,g}}{\partial W_{r}}\frac{\partial^{2}V^{\left(g\right)}}{\partial\tilde{p}_{\alpha,g}^{2}}\\
    &  & -\tau_{\alpha}\left(\frac{\partial p_{\alpha}}{\partial W_{s}}\frac{\partial V^{\left(r\right)}}{\partial\tilde{p}_{\alpha,r}}+\frac{\partial p_{\alpha}}{\partial W_{r}}\frac{\partial V^{\left(s\right)}}{\partial\tilde{p}_{\alpha,s}}\right)\\
    &  & +\frac{\partial^{2}p_{\alpha}}{\partial W_{s}\partial W_{r}}\sum_{g=1}^{G}\left[\frac{\partial V^{\left(g\right)}}{\partial\tilde{p}_{\alpha,g}}\left(1-\tau_{\alpha}W_{g}\right)\right].
\end{eqnarray*}
This gives the decomposition into three terms:

(1) $\sum_{g=1}^{G}\frac{\partial\tilde{p}_{\alpha,g}}{\partial W_{s}}\frac{\partial\tilde{p}_{\alpha,g}}{\partial W_{r}}\frac{\partial^{2}V^{\left(g\right)}}{\partial\tilde{p}_{\alpha,g}^{2}}:$ Note that
\begin{eqnarray*}
    \sum_{g=1}^{G}\frac{\partial\tilde{p}_{\alpha,g}}{\partial W_{s}}\frac{\partial\tilde{p}_{\alpha,g}}{\partial W_{r}}\frac{\partial^{2}V^{\left(g\right)}}{\partial\tilde{p}_{\alpha,g}^{2}} & = & \sum_{g=1}^{G}\left[\frac{\partial p_{\alpha}}{\partial W_{s}}\frac{\partial p_{\alpha}}{\partial W_{r}}\left(1-\tau_{\alpha}W_{g}\right)^{2}\frac{\partial^{2}V^{\left(g\right)}}{\partial\tilde{p}_{\alpha,g}^{2}}\right]\\
    &  & -\tau_{\alpha}p_{\alpha}\frac{\partial p_{\alpha}}{\partial W_{r}}\left(1-\tau_{\alpha}W_{s}\right)\frac{\partial^{2}V^{\left(s\right)}}{\partial\tilde{p}_{\alpha,s}^{2}}\\
    &  & -\tau_{\alpha}p_{\alpha}\frac{\partial p_{\alpha}}{\partial W_{s}}\left(1-\tau_{\alpha}W_{r}\right)\frac{\partial^{2}V^{\left(r\right)}}{\partial\tilde{p}_{\alpha,r}^{2}}\\
    & = & \frac{\partial p_{\alpha}}{\partial W_{s}}\frac{\partial p_{\alpha}}{\partial W_{r}}\sum_{g=1}^{G}\left(1-\tau_{\alpha}W_{g}\right)^{2}\frac{\partial^{2}V^{\left(g\right)}}{\partial\tilde{p}_{\alpha,g}^{2}}\\
    &  & -\tau_{\alpha}\left(\frac{\partial p_{\alpha}}{\partial W_{r}}\tilde{p}_{\alpha,s}\frac{\partial^{2}V^{\left(s\right)}}{\partial\tilde{p}_{\alpha,s}^{2}}+\frac{\partial p_{\alpha}}{\partial W_{s}}\tilde{p}_{\alpha,r}\frac{\partial^{2}V^{\left(r\right)}}{\partial\tilde{p}_{\alpha,r}^{2}}\right)
\end{eqnarray*}

(2) $-\tau_{\alpha}\left(\frac{\partial p_{\alpha}}{\partial W_{s}}\frac{\partial V^{\left(r\right)}}{\partial\tilde{p}_{\alpha,r}}+\frac{\partial p_{\alpha}}{\partial W_{r}}\frac{\partial V^{\left(s\right)}}{\partial\tilde{p}_{\alpha,s}}\right):$ By combining this with the last line of the preceding equations, we obtain

\begin{eqnarray*}
    &  & -\tau_{\alpha}\left(\frac{\partial p_{\alpha}}{\partial W_{r}}\tilde{p}_{\alpha,s}\frac{\partial^{2}V^{\left(s\right)}}{\partial\tilde{p}_{\alpha,s}^{2}}+\frac{\partial p_{\alpha}}{\partial W_{s}}\tilde{p}_{\alpha,r}\frac{\partial^{2}V^{\left(r\right)}}{\partial\tilde{p}_{\alpha,r}^{2}}\right)-\tau_{\alpha}\left(\frac{\partial p_{\alpha}}{\partial W_{s}}\frac{\partial V^{\left(r\right)}}{\partial\tilde{p}_{\alpha,r}}+\frac{\partial p_{\alpha}}{\partial W_{r}}\frac{\partial V^{\left(s\right)}}{\partial\tilde{p}_{\alpha,s}}\right)\\
    & = & -\tau_{\alpha}\left[\frac{\partial p_{\alpha}}{\partial W_{s}}\left(\frac{\partial V^{\left(r\right)}}{\partial\tilde{p}_{\alpha,r}}+\tilde{p}_{\alpha,r}\frac{\partial^{2}V^{\left(r\right)}}{\partial\tilde{p}_{\alpha,r}^{2}}\right)+\frac{\partial p_{\alpha}}{\partial W_{r}}\left(\frac{\partial V^{\left(s\right)}}{\partial\tilde{p}_{\alpha,s}}+\tilde{p}_{\alpha,s}\frac{\partial^{2}V^{\left(s\right)}}{\partial\tilde{p}_{\alpha,s}^{2}}\right)\right]\\
    & = & -\tau_{\alpha}\left[\frac{\partial p_{\alpha}}{\partial W_{s}}\left(\frac{\partial V^{\left(r\right)}}{\partial\tilde{p}_{\alpha,r}}-\frac{\partial V^{\left(r\right)}}{\partial\tilde{p}_{\alpha,r}}\frac{-\tilde{p}_{\alpha,r}\frac{\partial^{2}V^{\left(r\right)}}{\partial\tilde{p}_{\alpha,r}^{2}}}{\frac{\partial V^{\left(r\right)}}{\partial\tilde{p}_{\alpha,r}}}\right)+\frac{\partial p_{\alpha}}{\partial W_{r}}\left(\frac{\partial V^{\left(s\right)}}{\partial\tilde{p}_{\alpha,s}}-\frac{\partial V^{\left(s\right)}}{\partial\tilde{p}_{\alpha,s}}\frac{-\tilde{p}_{\alpha,s}\frac{\partial^{2}V^{\left(s\right)}}{\partial\tilde{p}_{\alpha,s}^{2}}}{\frac{\partial V^{\left(s\right)}}{\partial\tilde{p}_{\alpha,s}}}\right)\right]
\end{eqnarray*}

(3) $+\frac{\partial^{2}p_{\alpha}}{\partial W_{s}\partial W_{r}}\sum_{g=1}^{G}\left[\frac{\partial V^{\left(g\right)}}{\partial\tilde{p}_{\alpha,g}}\left(1-\tau_{\alpha}W_{g}\right)\right]:$ This corresponds to the third term. Hence, combining all terms, we obtain

\begin{eqnarray*}
    &  & \frac{\partial^{2}V}{\partial W_{s}\partial W_{r}}\\
    & = & \frac{\partial p_{\alpha}}{\partial W_{s}}\frac{\partial p_{\alpha}}{\partial W_{r}}\sum_{g=1}^{G}\left(1-\tau_{\alpha}W_{g}\right)^{2}\frac{\partial^{2}V^{\left(g\right)}}{\partial\tilde{p}_{\alpha,g}^{2}}\\
    &  & -\tau_{\alpha}\left[\frac{\partial p_{\alpha}}{\partial W_{s}}\left(1-\frac{-\tilde{p}_{\alpha,r}\frac{\partial^{2}V^{\left(r\right)}}{\partial\tilde{p}_{\alpha,r}^{2}}}{\frac{\partial V^{\left(r\right)}}{\partial\tilde{p}_{\alpha,r}}}\right)\frac{\partial V^{\left(r\right)}}{\partial\tilde{p}_{\alpha,r}}+\frac{\partial p_{\alpha}}{\partial W_{r}}\left(1-\frac{-\tilde{p}_{\alpha,s}\frac{\partial^{2}V^{\left(s\right)}}{\partial\tilde{p}_{\alpha,s}^{2}}}{\frac{\partial V^{\left(s\right)}}{\partial\tilde{p}_{\alpha,s}}}\right)\frac{\partial V^{\left(s\right)}}{\partial\tilde{p}_{\alpha,s}}\right]\\
    &  & +\frac{\partial^{2}p_{\alpha}}{\partial W_{s}\partial W_{r}}\sum_{g=1}^{G}\left[\frac{\partial V^{\left(g\right)}}{\partial\tilde{p}_{\alpha,g}}\left(1-\tau_{\alpha}W_{g}\right)\right].
\end{eqnarray*}
This establishes the second part of Lemma \ref{le2}.

\subsection{Proof of Lemma \ref{le3}}
From the market clearing condition
\begin{equation*}
    \sum_{g=1}^{G}Q_{\alpha}^{\left(g\right)}\left(\tilde{p}_{\alpha,g},p_{\beta}\right)=S_{\alpha}\left(p_{\alpha}\right),
\end{equation*}
differentiating both sides with respect to $W_{r}$ gives
\begin{eqnarray*}
    \sum_{g=1}^{G}\left[\frac{\partial\tilde{p}_{\alpha,g}}{\partial W_{r}}\frac{\partial Q_{\alpha}^{\left(g\right)}}{\partial\tilde{p}_{\alpha,g}}+\frac{\partial p_{\beta}}{\partial W_{r}}\frac{\partial Q_{\alpha}^{\left(g\right)}}{\partial p_{\beta}}\right] & = & \frac{\partial p_{\alpha}}{\partial W_{r}}\frac{\partial S_{\alpha}}{\partial p_{\alpha}}\\
    \sum_{g=1}^{G}\left[\left\{ \left(1-\tau_{\alpha}W_{g}\right)\frac{\partial p_{\alpha}}{\partial W_{r}}-\tau_{\alpha}p_{\alpha}\frac{\partial W_{g}}{\partial W_{r}}\right\} \frac{1}{1-\tau_{\alpha}W_{g}}\frac{\partial Q_{\alpha}^{\left(g\right)}}{\partial p_{\alpha}}+\frac{\partial p_{\beta}}{\partial W_{r}}\frac{\partial Q_{\alpha}^{\left(g\right)}}{\partial p_{\beta}}\right] & = & \frac{\partial p_{\alpha}}{\partial W_{r}}\frac{\partial S_{\alpha}}{\partial p_{\alpha}}\\
    -\tau_{\alpha}p_{\alpha}\frac{1}{1-\tau_{\alpha}W_{r}}\frac{\partial Q_{\alpha}^{\left(r\right)}}{\partial p_{\alpha}}+\frac{\partial p_{\alpha}}{\partial W_{r}}\sum_{g=1}^{G}\frac{\partial Q_{\alpha}^{\left(g\right)}}{\partial p_{\alpha}}+\frac{\partial p_{\beta}}{\partial W_{r}}\sum_{g=1}^{G}\frac{\partial Q_{\alpha}^{\left(g\right)}}{\partial p_{\beta}} & = & \frac{\partial p_{\alpha}}{\partial W_{r}}\frac{\partial S_{\alpha}}{\partial p_{\alpha}}\\
    \frac{\partial p_{\alpha}}{\partial W_{r}}\left\{ \sum_{g=1}^{G}\frac{\partial Q_{\alpha}^{\left(g\right)}}{\partial p_{\alpha}}-\frac{\partial S_{\alpha}}{\partial p_{\alpha}}\right\} +\frac{\partial p_{\beta}}{\partial W_{r}}\sum_{g=1}^{G}\frac{\partial Q_{\alpha}^{\left(g\right)}}{\partial p_{\beta}} & = & \frac{\tau_{\alpha}p_{\alpha}}{1-\tau_{\alpha}W_{r}}\frac{\partial Q_{\alpha}^{\left(r\right)}}{\partial p_{\alpha}}
\end{eqnarray*}
Similarly, from the condition
\begin{equation*}
    \sum_{g=1}^{G}Q_{\beta}^{\left(g\right)}\left(\tilde{p}_{\alpha,g},p_{\beta}\right)=S_{\beta}\left(p_{\beta}\right),
\end{equation*}
differentiating both sides with respect to $W_{r}$ gives
\begin{eqnarray*}
    \sum_{g=1}^{G}\left[\frac{\partial\tilde{p}_{\alpha,g}}{\partial W_{r}}\frac{\partial Q_{\beta}^{\left(g\right)}}{\partial\tilde{p}_{\alpha,g}}+\frac{\partial p_{\beta}}{\partial W_{r}}\frac{\partial Q_{\beta}^{\left(g\right)}}{\partial p_{\beta}}\right] & = & \frac{\partial p_{\beta}}{\partial W_{r}}\frac{\partial S_{\beta}}{\partial p_{\beta}}\\
    \sum_{g=1}^{G}\left[\left\{ \left(1-\tau_{\alpha}W_{g}\right)\frac{\partial p_{\alpha}}{\partial W_{r}}-\tau_{\alpha}p_{\alpha}\frac{\partial W_{g}}{\partial W_{r}}\right\} \frac{1}{1-\tau_{\alpha}W_{g}}\frac{\partial Q_{\beta}^{\left(g\right)}}{\partial p_{\alpha}}+\frac{\partial p_{\beta}}{\partial W_{r}}\frac{\partial Q_{\beta}^{\left(g\right)}}{\partial p_{\beta}}\right] & = & \frac{\partial p_{\beta}}{\partial W_{r}}\frac{\partial S_{\beta}}{\partial p_{\beta}}\\
    -\tau_{\alpha}p_{\alpha}\frac{1}{1-\tau_{\alpha}W_{r}}\frac{\partial Q_{\beta}^{\left(r\right)}}{\partial p_{\alpha}}+\frac{\partial p_{\alpha}}{\partial W_{r}}\sum_{g=1}^{G}\frac{\partial Q_{\beta}^{\left(g\right)}}{\partial p_{\alpha}}+\frac{\partial p_{\beta}}{\partial W_{r}}\sum_{g=1}^{G}\frac{\partial Q_{\beta}^{\left(g\right)}}{\partial p_{\beta}} & = & \frac{\partial p_{\beta}}{\partial W_{r}}\frac{\partial S_{\beta}}{\partial p_{\beta}}\\
    \frac{\partial p_{\alpha}}{\partial W_{r}}\sum_{g=1}^{G}\frac{\partial Q_{\beta}^{\left(g\right)}}{\partial p_{\alpha}}+\frac{\partial p_{\beta}}{\partial W_{r}}\left\{ \sum_{g=1}^{G}\frac{\partial Q_{\beta}^{\left(g\right)}}{\partial p_{\beta}}-\frac{\partial S_{\beta}}{\partial p_{\beta}}\right\}  & = & \frac{\tau_{\alpha}p_{\alpha}}{1-\tau_{\alpha}W_{r}}\frac{\partial Q_{\beta}^{\left(r\right)}}{\partial p_{\alpha}}
\end{eqnarray*}
These two conditions can be written as
\begin{equation*}
    \left[\begin{array}{cc}
\sum_{g=1}^{G}\frac{\partial Q_{\alpha}^{\left(g\right)}}{\partial p_{\alpha}}-\frac{\partial S_{\alpha}}{\partial p_{\alpha}} & \sum_{g=1}^{G}\frac{\partial Q_{\alpha}^{\left(g\right)}}{\partial p_{\beta}}\\
\sum_{g=1}^{G}\frac{\partial Q_{\beta}^{\left(g\right)}}{\partial p_{\alpha}} & \sum_{g=1}^{G}\frac{\partial Q_{\beta}^{\left(g\right)}}{\partial p_{\beta}}-\frac{\partial S_{\beta}}{\partial p_{\beta}}
\end{array}\right]\left[\begin{array}{c}
\frac{\partial p_{\alpha}}{\partial W_{r}}\\
\frac{\partial p_{\beta}}{\partial W_{r}}
\end{array}\right]=\frac{\tau_{\alpha}p_{\alpha}}{1-\tau_{\alpha}W_{r}}\left[\begin{array}{c}
\frac{\partial Q_{\alpha}^{\left(r\right)}}{\partial p_{\alpha}}\\
\frac{\partial Q_{\beta}^{\left(r\right)}}{\partial p_{\alpha}}
\end{array}\right].
\end{equation*}
Therefore,
\begin{eqnarray*}
    \left[\begin{array}{c}
\frac{\partial p_{\alpha}}{\partial W_{r}}\\
\frac{\partial p_{\beta}}{\partial W_{r}}
\end{array}\right] & = & \frac{\tau_{\alpha}p_{\alpha}}{1-\tau_{\alpha}W_{r}}\left[\begin{array}{cc}
\sum_{g=1}^{G}\frac{\partial Q_{\alpha}^{\left(g\right)}}{\partial p_{\alpha}}-\frac{\partial S_{\alpha}}{\partial p_{\alpha}} & \sum_{g=1}^{G}\frac{\partial Q_{\alpha}^{\left(g\right)}}{\partial p_{\beta}}\\
\sum_{g=1}^{G}\frac{\partial Q_{\beta}^{\left(g\right)}}{\partial p_{\alpha}} & \sum_{g=1}^{G}\frac{\partial Q_{\beta}^{\left(g\right)}}{\partial p_{\beta}}-\frac{\partial S_{\beta}}{\partial p_{\beta}}
\end{array}\right]^{-1}\left[\begin{array}{c}
\frac{\partial Q_{\alpha}^{\left(r\right)}}{\partial p_{\alpha}}\\
\frac{\partial Q_{\beta}^{\left(r\right)}}{\partial p_{\alpha}}
\end{array}\right]\\
    & = & \frac{\frac{\tau_{\alpha}p_{\alpha}}{1-\tau_{\alpha}W_{r}}}{\left(\sum_{g=1}^{G}\frac{\partial Q_{\alpha}^{\left(g\right)}}{\partial p_{\alpha}}-\frac{\partial S_{\alpha}}{\partial p_{\alpha}}\right)\left(\sum_{g=1}^{G}\frac{\partial Q_{\beta}^{\left(g\right)}}{\partial p_{\beta}}-\frac{\partial S_{\beta}}{\partial p_{\beta}}\right)-\left(\sum_{g=1}^{G}\frac{\partial Q_{\alpha}^{\left(g\right)}}{\partial p_{\beta}}\right)\left(\sum_{g=1}^{G}\frac{\partial Q_{\beta}^{\left(g\right)}}{\partial p_{\alpha}}\right)}\\
    &  & \times\left[\begin{array}{cc}
\sum_{g=1}^{G}\frac{\partial Q_{\beta}^{\left(g\right)}}{\partial p_{\beta}}-\frac{\partial S_{\beta}}{\partial p_{\beta}} & -\sum_{g=1}^{G}\frac{\partial Q_{\alpha}^{\left(g\right)}}{\partial p_{\beta}}\\
-\sum_{g=1}^{G}\frac{\partial Q_{\beta}^{\left(g\right)}}{\partial p_{\alpha}} & \sum_{g=1}^{G}\frac{\partial Q_{\alpha}^{\left(g\right)}}{\partial p_{\alpha}}-\frac{\partial S_{\alpha}}{\partial p_{\alpha}}
\end{array}\right]\left[\begin{array}{c}
\frac{\partial Q_{\alpha}^{\left(r\right)}}{\partial p_{\alpha}}\\
\frac{\partial Q_{\beta}^{\left(r\right)}}{\partial p_{\alpha}}
\end{array}\right].
\end{eqnarray*}
In particular,
\begin{equation*}
    \frac{\partial p_{\alpha}}{\partial W_{r}}=\frac{\frac{\tau_{\alpha}p_{\alpha}}{1-\tau_{\alpha}W_{r}}\left[\left(\sum_{g=1}^{G}\frac{\partial Q_{\beta}^{\left(g\right)}}{\partial p_{\beta}}-\frac{\partial S_{\beta}}{\partial p_{\beta}}\right)\frac{\partial Q_{\alpha}^{\left(r\right)}}{\partial p_{\alpha}}-\left(\sum_{g=1}^{G}\frac{\partial Q_{\alpha}^{\left(g\right)}}{\partial p_{\beta}}\right)\frac{\partial Q_{\beta}^{\left(r\right)}}{\partial p_{\alpha}}\right]}{\left(\sum_{g=1}^{G}\frac{\partial Q_{\alpha}^{\left(g\right)}}{\partial p_{\alpha}}-\frac{\partial S_{\alpha}}{\partial p_{\alpha}}\right)\left(\sum_{g=1}^{G}\frac{\partial Q_{\beta}^{\left(g\right)}}{\partial p_{\beta}}-\frac{\partial S_{\beta}}{\partial p_{\beta}}\right)-\left(\sum_{g=1}^{G}\frac{\partial Q_{\alpha}^{\left(g\right)}}{\partial p_{\beta}}\right)\left(\sum_{g=1}^{G}\frac{\partial Q_{\beta}^{\left(g\right)}}{\partial p_{\alpha}}\right)}.
\end{equation*}
The denominator is
\begin{eqnarray*}
    &  & \left(\sum_{g=1}^{G}\frac{\partial Q_{\alpha}^{\left(g\right)}}{\partial p_{\alpha}}-\frac{\partial S_{\alpha}}{\partial p_{\alpha}}\right)\left(\sum_{g=1}^{G}\frac{\partial Q_{\beta}^{\left(g\right)}}{\partial p_{\beta}}-\frac{\partial S_{\beta}}{\partial p_{\beta}}\right)\\
    & > & \left(\sum_{g=1}^{G}\frac{\partial Q_{\alpha}^{\left(g\right)}}{\partial p_{\alpha}}\right)\left(\sum_{g=1}^{G}\frac{\partial Q_{\beta}^{\left(g\right)}}{\partial p_{\beta}}\right)\\
    & = & \left|\sum_{g=1}^{G}\frac{\partial Q_{\alpha}^{\left(g\right)}}{\partial p_{\alpha}}\right|\left|\sum_{g=1}^{G}\frac{\partial Q_{\beta}^{\left(g\right)}}{\partial p_{\beta}}\right|\\
    & > & \left|\sum_{g=1}^{G}\frac{\partial Q_{\alpha}^{\left(g\right)}}{\partial p_{\beta}}\right|\left|\sum_{g=1}^{G}\frac{\partial Q_{\beta}^{\left(g\right)}}{\partial p_{\alpha}}\right|\\
    & \geq & \left(\sum_{g=1}^{G}\frac{\partial Q_{\alpha}^{\left(g\right)}}{\partial p_{\beta}}\right)\left(\sum_{g=1}^{G}\frac{\partial Q_{\beta}^{\left(g\right)}}{\partial p_{\alpha}}\right).
\end{eqnarray*}
By Assumption \ref{as3}, own-price effects dominate cross-price effects, so the inequality in the second-to-last line holds strictly. Hence the denominator is positive. Similarly, the numerator is
\begin{eqnarray*}
    &  & \left(\sum_{g=1}^{G}\frac{\partial Q_{\beta}^{\left(g\right)}}{\partial p_{\beta}}-\frac{\partial S_{\beta}}{\partial p_{\beta}}\right)\frac{\partial Q_{\alpha}^{\left(r\right)}}{\partial p_{\alpha}}\\
    & > & \left(\sum_{g=1}^{G}\frac{\partial Q_{\beta}^{\left(g\right)}}{\partial p_{\beta}}\right)\frac{\partial Q_{\alpha}^{\left(r\right)}}{\partial p_{\alpha}}\\
    & = & \left|\sum_{g=1}^{G}\frac{\partial Q_{\beta}^{\left(g\right)}}{\partial p_{\beta}}\right|\left|\frac{\partial Q_{\alpha}^{\left(r\right)}}{\partial p_{\alpha}}\right|\\
    & > & \left|\sum_{g=1}^{G}\frac{\partial Q_{\alpha}^{\left(g\right)}}{\partial p_{\beta}}\right|\left|\frac{\partial Q_{\beta}^{\left(r\right)}}{\partial p_{\alpha}}\right|\\
    & \geq & \left(\sum_{g=1}^{G}\frac{\partial Q_{\alpha}^{\left(g\right)}}{\partial p_{\beta}}\right)\frac{\partial Q_{\beta}^{\left(r\right)}}{\partial p_{\alpha}}.
\end{eqnarray*}
By Assumption \ref{as3}, the inequality in the second-to-last line holds strictly. Hence the numerator is positive. Since both the numerator and denominator are positive, it follows that
\begin{equation*}
    \frac{\partial p_{\alpha}}{\partial W_{r}}>0.
\end{equation*}

\subsection{Proof of Theorem \ref{th4}}
Recall that
\begin{eqnarray*}
    &  & \frac{\partial^{2}V}{\partial W_{s}\partial W_{r}}\\
    & = & \frac{\partial p_{\alpha}}{\partial W_{s}}\frac{\partial p_{\alpha}}{\partial W_{r}}\sum_{g=1}^{G}\left(1-\tau_{\alpha}W_{g}\right)^{2}\frac{\partial^{2}V^{\left(g\right)}}{\partial\tilde{p}_{\alpha,g}^{2}}\\
    &  & -\tau_{\alpha}\left[\frac{\partial p_{\alpha}}{\partial W_{s}}\left(1-\frac{-\tilde{p}_{\alpha,r}\frac{\partial^{2}V^{\left(r\right)}}{\partial\tilde{p}_{\alpha,r}^{2}}}{\frac{\partial V^{\left(r\right)}}{\partial\tilde{p}_{\alpha,r}}}\right)\frac{\partial V^{\left(r\right)}}{\partial\tilde{p}_{\alpha,r}}+\frac{\partial p_{\alpha}}{\partial W_{r}}\left(1-\frac{-\tilde{p}_{\alpha,s}\frac{\partial^{2}V^{\left(s\right)}}{\partial\tilde{p}_{\alpha,s}^{2}}}{\frac{\partial V^{\left(s\right)}}{\partial\tilde{p}_{\alpha,s}}}\right)\frac{\partial V^{\left(s\right)}}{\partial\tilde{p}_{\alpha,s}}\right]\\
    &  & +\frac{\partial^{2}p_{\alpha}}{\partial W_{s}\partial W_{r}}\sum_{g=1}^{G}\left[\frac{\partial V^{\left(g\right)}}{\partial\tilde{p}_{\alpha,g}}\left(1-\tau_{\alpha}W_{g}\right)\right].
\end{eqnarray*}
The first line is positive under Assumptions \ref{as2} and \ref{as3}. The second line is positive under Assumptions \ref{as2}, \ref{as3}, and \ref{as4}. The third line is positive under Assumptions \ref{as2} and \ref{as5}. Therefore, the value function is supermodular in $\boldsymbol{W}$ under these assumptions.

\subsection{Proof of Theorem \ref{th5}}
Let $V^{*}\equiv V\left(\boldsymbol{W}^{*};\theta^{*}\right)$ and $\hat{V}\equiv V\left(\hat{\boldsymbol{W}};\hat{\theta}\right).$ We can decompose the welfare risk $\left|\mathcal{R}\right|=\left|V^{*}-\hat{V}\right|$ as follows. First, note that
\begin{eqnarray*}
    V^{*}-\hat{V} & = & V\left(\boldsymbol{W}^{*};\theta^{*}\right)-V\left(\hat{\boldsymbol{W}};\hat{\theta}\right)\\
    & = & \left\{ V\left(\boldsymbol{W}^{*};\theta^{*}\right)-V\left(\boldsymbol{W}^{*};\hat{\theta}\right)\right\} +\left\{ V\left(\boldsymbol{W}^{*};\hat{\theta}\right)-V\left(\hat{\boldsymbol{W}};\hat{\theta}\right)\right\} \\
    & \leq & V\left(\boldsymbol{W}^{*};\theta^{*}\right)-V\left(\boldsymbol{W}^{*};\hat{\theta}\right)
\end{eqnarray*}
where the final line holds because $\hat{\boldsymbol{W}}$ is optimal for $\hat{\theta}$ so that $V\left(\boldsymbol{W}^{*};\hat{\theta}\right)-V\left(\hat{\boldsymbol{W}};\hat{\theta}\right)\leq0$. So, we obtain
\begin{eqnarray*}
    V^{*}-\hat{V} & \leq & V\left(\boldsymbol{W}^{*};\theta^{*}\right)-V\left(\boldsymbol{W}^{*};\hat{\theta}\right)\\
    & \leq & \left|V\left(\boldsymbol{W}^{*};\theta^{*}\right)-V\left(\boldsymbol{W}^{*};\hat{\theta}\right)\right|\\
    & \leq & \sup_{\boldsymbol{W}}\left|V\left(\boldsymbol{W};\theta^{*}\right)-V\left(\boldsymbol{W};\hat{\theta}\right)\right|.
\end{eqnarray*}
Similarly,
\begin{eqnarray*}
    \hat{V}-V^{*} & = & V\left(\hat{\boldsymbol{W}};\hat{\theta}\right)-V\left(\boldsymbol{W}^{*};\theta^{*}\right)\\
    & = & \left\{ V\left(\hat{\boldsymbol{W}};\hat{\theta}\right)-V\left(\hat{\boldsymbol{W}};\theta^{*}\right)\right\} +\left\{ V\left(\hat{\boldsymbol{W}};\theta^{*}\right)-V\left(\boldsymbol{W}^{*};\theta^{*}\right)\right\} \\
    & \leq & V\left(\hat{\boldsymbol{W}};\hat{\theta}\right)-V\left(\hat{\boldsymbol{W}};\theta^{*}\right)
\end{eqnarray*}
where the final line holds because $\boldsymbol{W}^{*}$ is optimal for $\theta^{*}$ so that $V\left(\hat{\boldsymbol{W}};\theta^{*}\right)-V\left(\boldsymbol{W}^{*};\theta^{*}\right)\leq0$. So, we obtain
\begin{eqnarray*}
    \hat{V}-V^{*} & \leq & V\left(\hat{\boldsymbol{W}};\hat{\theta}\right)-V\left(\hat{\boldsymbol{W}};\theta^{*}\right)\\
    & \leq & \left|V\left(\hat{\boldsymbol{W}};\hat{\theta}\right)-V\left(\hat{\boldsymbol{W}};\theta^{*}\right)\right|\\
    & \leq & \sup_{\boldsymbol{W}}\left|V\left(\boldsymbol{W};\hat{\theta}\right)-V\left(\boldsymbol{W};\theta^{*}\right)\right|.
\end{eqnarray*}
Combining the two inequalities gives
\begin{eqnarray*}
    \left|\mathcal{R}\right| & = & \left|\hat{V}-V^{*}\right|\\
    & \leq & \sup_{\boldsymbol{W}}\left|V\left(\boldsymbol{W};\hat{\theta}\right)-V\left(\boldsymbol{W};\theta^{*}\right)\right|\\
    & \leq & \overline{S}\left\Vert \hat{\theta}-\theta^{*}\right\Vert 
\end{eqnarray*}
for a constant $\overline{S}>0.$ The final inequality follows because the planner's objective function is Lipschitz continuous in $\theta,$ as it is a linear combination of each group's indirect utility function under Assumption \ref{as6}. Taking expectation gives
\begin{equation*}
    \mathbb{E}\left|\mathcal{R}\right|\leq\overline{S}\mathbb{E}\left\Vert \hat{\theta}-\theta^{*}\right\Vert .
\end{equation*}
Since $\left\Vert \hat{\theta}-\theta^{*}\right\Vert $ is a nonnegative random variable, we may write
\begin{equation*}
    \mathbb{E}\left\Vert \hat{\theta}-\theta^{*}\right\Vert =\int_{0}^{\infty}\mathrm{Pr}\left(\left\Vert \hat{\theta}-\theta^{*}\right\Vert >u\right)du.
\end{equation*}
By applying Assumption \ref{as7}, the expectation can be bounded as
\begin{eqnarray*}
    \mathbb{E}\left\Vert \hat{\theta}-\theta^{*}\right\Vert  & = & \int_{0}^{\infty}\mathrm{Pr}\left(\left\Vert \hat{\theta}-\theta^{*}\right\Vert >u\right)du\\
    & \leq & \int_{0}^{\infty}c_{1}\exp\left(-c_{2}nu^{2}\right)du\\
    & = & \frac{c_{1}}{\sqrt{n}}\int_{0}^{\infty}\exp\left(-c_{2}z^{2}\right)dz\quad(z=\sqrt{n}u)\\
    & = & \frac{c_{1}}{\sqrt{n}}\frac{1}{2}\sqrt{\frac{\pi}{c_{2}}}.
\end{eqnarray*}
Thus,
\begin{eqnarray*}
    \mathbb{E}\left|\mathcal{R}\right| & \leq & \overline{S}\mathbb{E}\left\Vert \hat{\theta}-\theta^{*}\right\Vert \\
    & \leq & \frac{\overline{S}}{\sqrt{n}}\frac{c_{1}}{2}\sqrt{\frac{\pi}{c_{2}}},    
\end{eqnarray*}
which yields the desired result.

\subsection{Proof of Example \ref{ex1}}
Note that
\begin{eqnarray*}
    &  & V\left(1,1\right)\\
    & = & -\left(\gamma_{1}+\gamma_{2}\right)\log\left(1-\tau_{\alpha}\right)-\frac{\log\left(\frac{\gamma_{1}M_{1}}{1-\tau_{\alpha}}+\frac{\gamma_{2}M_{2}}{1-\tau_{\alpha}}\right)}{1+\kappa_{\alpha}}\left(\gamma_{1}+\gamma_{2}\right)\\
    & = & -\left(\gamma_{1}+\gamma_{2}\right)\log\left(1-\tau_{\alpha}\right)-\frac{\log\left(\gamma_{1}M_{1}+\gamma_{2}M_{2}\right)-\log\left(1-\tau_{\alpha}\right)}{1+\kappa_{\alpha}}\left(\gamma_{1}+\gamma_{2}\right)\\
    & = & -\left(\gamma_{1}+\gamma_{2}\right)\log\left(1-\tau_{\alpha}\right)+\frac{\log\left(1-\tau_{\alpha}\right)}{1+\kappa_{\alpha}}\left(\gamma_{1}+\gamma_{2}\right)-\frac{\log\left(\gamma_{1}M_{1}+\gamma_{2}M_{2}\right)}{1+\kappa_{\alpha}}\left(\gamma_{1}+\gamma_{2}\right)\\
    & = & -\left\{ 1-\frac{1}{1+\kappa_{\alpha}}\right\} \left(\gamma_{1}+\gamma_{2}\right)\log\left(1-\tau_{\alpha}\right)-\frac{\log\left(\gamma_{1}M_{1}+\gamma_{2}M_{2}\right)}{1+\kappa_{\alpha}}\left(\gamma_{1}+\gamma_{2}\right)\\
    & = & -\frac{\kappa_{\alpha}}{1+\kappa_{\alpha}}\left(\gamma_{1}+\gamma_{2}\right)\log\left(1-\tau_{\alpha}\right)+V^{*}\left(0,0\right).
\end{eqnarray*}
Thus,
\begin{eqnarray*}
    V\left(1,1\right)-V\left(0,0\right) & = & -\frac{\kappa_{\alpha}}{1+\kappa_{\alpha}}\left(\gamma_{1}+\gamma_{2}\right)\log\left(1-\tau_{\alpha}\right)\\
    & > & 0.
\end{eqnarray*}

\subsection{Proof of Example \ref{ex2}}
Note that
\begin{eqnarray*}
     &  & V\left(1,0\right)-V\left(0,1\right)\\
     & = & \left[-\gamma_{1}\log\left(1-\tau_{\alpha}\right)-\frac{\log\left(\frac{\gamma_{1}M_{1}}{1-\tau_{\alpha}}+\gamma_{2}M_{2}\right)}{1+\kappa_{\alpha}}\left(\gamma_{1}+\gamma_{2}\right)\right]\\
     &  & -\left[-\gamma_{2}\log\left(1-\tau_{\alpha}\right)-\frac{\log\left(\gamma_{1}M_{1}+\frac{\gamma_{2}M_{2}}{1-\tau_{\alpha}}\right)}{1+\kappa_{\alpha}}\left(\gamma_{1}+\gamma_{2}\right)\right]\\
     & = & \frac{2\gamma}{1+\kappa_{\alpha}}\left[-\log\left(\frac{\gamma M_{1}}{1-\tau_{\alpha}}+\gamma M_{2}\right)+\log\left(\gamma M_{1}+\frac{\gamma M_{2}}{1-\tau_{\alpha}}\right)\right]\\
     & = & \frac{2\gamma}{1+\kappa_{\alpha}}\log\left(\frac{\gamma M_{1}+\frac{\gamma M_{2}}{1-\tau_{\alpha}}}{\frac{\gamma M_{1}}{1-\tau_{\alpha}}+\gamma M_{2}}\right)\\
     & = & \frac{2\gamma}{1+\kappa_{\alpha}}\log\left(\frac{M_{1}\left(1-\tau_{\alpha}\right)+M_{2}}{M_{1}+M_{2}\left(1-\tau_{\alpha}\right)}\right)
\end{eqnarray*}
and we see that
\begin{equation*}
    \left[M_{1}\left(1-\tau_{\alpha}\right)+M_{2}\right]-\left[M_{1}+M_{2}\left(1-\tau_{\alpha}\right)\right]=\tau_{\alpha}\left(M_{2}-M_{1}\right).
\end{equation*}
Thus, the sign of $V\left(1,0\right)-V\left(0,1\right)$ is equivalent to that of $M_{2}-M_{1}.$

\subsection{Proof of Example \ref{ex3}}
Note that
\begin{eqnarray*}
    &  & V\left(1,0\right)-V\left(0,1\right)\\
    & = & \left[-\gamma_{1}\log\left(1-\tau_{\alpha}\right)-\frac{\log\left(\frac{\gamma_{1}M_{1}}{1-\tau_{\alpha}}+\gamma_{2}M_{2}\right)}{1+\kappa_{\alpha}}\left(\gamma_{1}+\gamma_{2}\right)\right]\\
    &  & -\left[-\gamma_{2}\log\left(1-\tau_{\alpha}\right)-\frac{\log\left(\gamma_{1}M_{1}+\frac{\gamma_{2}M_{2}}{1-\tau_{\alpha}}\right)}{1+\kappa_{\alpha}}\left(\gamma_{1}+\gamma_{2}\right)\right]\\
    & = & \left(\gamma_{2}-\gamma_{1}\right)\log\left(1-\tau_{\alpha}\right)+\frac{\left(\gamma_{1}+\gamma_{2}\right)}{1+\kappa_{\alpha}}\left\{ \log\left(\gamma_{1}M+\frac{\gamma_{2}M}{1-\tau_{\alpha}}\right)-\log\left(\frac{\gamma_{1}M}{1-\tau_{\alpha}}+\gamma_{2}M\right)\right\} \\
    & = & \left(\gamma_{2}-\gamma_{1}\right)\log\left(1-\tau_{\alpha}\right)+\frac{\left(\gamma_{1}+\gamma_{2}\right)}{1+\kappa_{\alpha}}\log\left(\frac{\gamma_{1}\left(1-\tau_{\alpha}\right)+\gamma_{2}}{\gamma_{1}+\gamma_{2}\left(1-\tau_{\alpha}\right)}\right)\\
    & = & \left(\gamma_{2}-\gamma_{1}\right)\log\left(1-\tau_{\alpha}\right)+\frac{\left(\gamma_{1}+\gamma_{2}\right)}{1+\kappa_{\alpha}}\log\left(1+\frac{\tau\left(\gamma_{2}-\gamma_{1}\right)}{\gamma_{1}+\gamma_{2}\left(1-\tau_{\alpha}\right)}\right)
\end{eqnarray*}
and the sign of it is equivalent to that of $\gamma_{1}-\gamma_{2}.$

\subsection{Proof of Example \ref{ex4}}
Note that
\begin{eqnarray*}
    &  & V\left(1,1\right)-V\left(1,0\right)\\
    & = & \left[-\left(\gamma_{1}+\gamma_{2}\right)\log\left(1-\tau_{\alpha}\right)-\frac{\log\left(\frac{\gamma_{1}M_{1}}{1-\tau_{\alpha}}+\frac{\gamma_{2}M_{2}}{1-\tau_{\alpha}}\right)}{1+\kappa_{\alpha}}\left(\gamma_{1}+\gamma_{2}\right)\right]\\
    &  & -\left[-\gamma_{1}\log\left(1-\tau_{\alpha}\right)-\frac{\log\left(\frac{\gamma_{1}M_{1}}{1-\tau_{\alpha}}+\gamma_{2}M_{2}\right)}{1+\kappa_{\alpha}}\left(\gamma_{1}+\gamma_{2}\right)\right]\\
    & = & -\gamma\log\left(1-\tau_{\alpha}\right)-\frac{2\gamma}{1+\kappa_{\alpha}}\left[\log\left(\frac{\gamma M_{1}}{1-\tau_{\alpha}}+\frac{\gamma M_{2}}{1-\tau_{\alpha}}\right)-\log\left(\frac{\gamma M_{1}}{1-\tau_{\alpha}}+\gamma M_{2}\right)\right]\\
    & = & -\gamma\log\left(1-\tau_{\alpha}\right)-\frac{2\gamma}{1+\kappa_{\alpha}}\log\left(\frac{\frac{M_{1}}{1-\tau_{\alpha}}+\frac{M_{2}}{1-\tau_{\alpha}}}{\frac{M_{1}}{1-\tau_{\alpha}}+M_{2}}\right)\\
    & = & -\gamma\log\left(1-\tau_{\alpha}\right)-\frac{2\gamma}{1+\kappa_{\alpha}}\log\left(\frac{M_{1}+M_{2}}{M_{1}+M_{2}\left(1-\tau_{\alpha}\right)}\right)\\
    & \geq & -\gamma\log\left(1-\tau_{\alpha}\right)-\gamma\log\left(\frac{M_{1}+M_{2}}{M_{1}+M_{2}\left(1-\tau_{\alpha}\right)}\right)\\
    & = & -\gamma\log\left(1-\tau_{\alpha}\right)+\gamma\log\left(\frac{M_{1}+M_{2}\left(1-\tau_{\alpha}\right)}{M_{1}+M_{2}}\right)\\
    & = & -\gamma\log\left(1-\tau_{\alpha}\right)+\gamma\log\left(1-\tau_{\alpha}\frac{M_{2}}{M_{1}+M_{2}}\right)\\    
    & > & -\gamma\log\left(1-\tau_{\alpha}\right)+\gamma\log\left(1-\tau_{\alpha}\right)\\
    & = & 0.
\end{eqnarray*}
Similar derivation gives $V\left(1,1\right)-V\left(0,1\right)>0.$

\subsection{Proof of Example \ref{ex5}}
Note that from Example \ref{ex4}, we have
\begin{eqnarray*}
    V\left(1,1\right)-V\left(1,0\right) & = & -\gamma\log\left(1-\tau_{\alpha}\right)-\frac{2\gamma}{1+\kappa_{\alpha}}\log\left(\frac{M_{1}+M_{2}}{M_{1}+M_{2}\left(1-\tau_{\alpha}\right)}\right)\\
    & = & -\gamma\log\left(1-\tau_{\alpha}\right)+\frac{2\gamma}{1+\kappa_{\alpha}}\log\left(\frac{M_{1}+M_{2}\left(1-\tau_{\alpha}\right)}{M_{1}+M_{2}}\right)\\
    & = & \gamma\left[\frac{2}{1+\kappa_{\alpha}}\log\left(\frac{M_{1}+M_{2}\left(1-\tau_{\alpha}\right)}{M_{1}+M_{2}}\right)-\log\left(1-\tau_{\alpha}\right)\right]\\
    & = & \frac{2\gamma}{1+\kappa_{\alpha}}\left[\log\left(\frac{M_{1}+M_{2}\left(1-\tau_{\alpha}\right)}{M_{1}+M_{2}}\right)-\log\left(1-\tau_{\alpha}\right)^{\frac{1+\kappa_{\alpha}}{2}}\right].
\end{eqnarray*}
Define the income ratio $M_{12}\equiv\frac{M_{1}}{M_{2}}.$ Then, $V\left(1,1\right)<V\left(1,0\right)$ if and only if
\begin{equation*}
    \frac{M_{1}+M_{2}\left(1-\tau_{\alpha}\right)}{M_{1}+M_{2}}<\left(1-\tau_{\alpha}\right)^{\frac{1+\kappa_{\alpha}}{2}},
\end{equation*}
which is equivalent to
\begin{equation*}
    M_{1}+M_{2}\left(1-\tau_{\alpha}\right)<\left(1-\tau_{\alpha}\right)^{\frac{1+\kappa_{\alpha}}{2}}\left(M_{1}+M_{2}\right).
\end{equation*}
Divide both sides by $M_{2}$ so that the above condition becomes
\begin{equation*}
    M_{12}+\left(1-\tau_{\alpha}\right)<\left(1-\tau_{\alpha}\right)^{\frac{1+\kappa_{\alpha}}{2}}\left(M_{12}+1\right)
\end{equation*}
which is equivalent to
\begin{equation*}
    M_{12}\left\{ 1-\left(1-\tau_{\alpha}\right)^{\frac{1+\kappa_{\alpha}}{2}}\right\} <\left(1-\tau_{\alpha}\right)^{\frac{1+\kappa_{\alpha}}{2}}-\left(1-\tau_{\alpha}\right)
\end{equation*}
which is equivalent to
\begin{equation*}
    M_{12}<\frac{\left(1-\tau_{\alpha}\right)^{\frac{1+\kappa_{\alpha}}{2}}-\left(1-\tau_{\alpha}\right)}{1-\left(1-\tau_{\alpha}\right)^{\frac{1+\kappa_{\alpha}}{2}}}\approx\frac{1-\kappa_{\alpha}}{1+\kappa_{\alpha}}
\end{equation*}
where the final line is computed by the binomial approximation.

\bibliographystyle{apalike} 
\bibliography{bibliography}

\end{document}